\documentclass[%
 aip,
 amsmath,amssymb,
preprint,%
]{revtex4-1}

\usepackage{graphicx}
\usepackage{dcolumn}
\usepackage{bm}

\usepackage[utf8]{inputenc}
\usepackage[T1]{fontenc}
\usepackage{mathptmx}
\date{}

\usepackage{enumitem}
\usepackage{gensymb}
\usepackage{amsmath}
\usepackage{amssymb}
\usepackage{color}
\usepackage{verbatim}

\newcommand{\Var}{\textrm{Var}}

\newcommand{\R}{ {\bf R} }
\newcommand{\br}{ {\bf r} }
\newcommand{\PsiT}{ \Psi_T }
\newcommand{\PsiTR}{ \Psi_T({\bf R}) }
\newcommand{\PsiTRp}{ \Psi_T({\bf R'}) }

\newcommand{\fRt}{ f({\bf R},t) }
\newcommand{\VR}{ {\bf V(R)} }
\newcommand{\D}{ {\cal D} }
\newcommand{\DR}{ {\cal D}({\bf R}) }
\newcommand{\J}{ {\cal J} }

\newcommand{\eJ}{ \exp{\cal J} }
\newcommand{\eJR}{ \exp{\cal J}({\bf R}) }

\newcommand{\bra}[1]{\left< #1 \right|}
\newcommand{\ket}[1]{\left| #1 \right>}
\newcommand{\braket}[2]{\left< #1 \middle| #2 \right>}

\newcommand{\revision}[1]{#1}

\begin{document}

\title{\bf A new scheme for fixed node diffusion quantum Monte Carlo with pseudopotentials: improving reproducibility and reducing the trial-wave-function bias}

\author{Andrea Zen} 
\email{a.zen@ucl.ac.uk}
\affiliation{ Thomas Young Centre, University College London, London WC1H 0AH, UK. }
\affiliation{ London Centre for Nanotechnology, Gordon St., London WC1H 0AH, UK. }
\affiliation{ Dept. of Physics and Astronomy, University College London, London WC1E 6BT, UK. }
\affiliation{ Dept. of Earth Sciences, University College London, London WC1E 6BT, UK. }

\author{Jan Gerit Brandenburg} 
\affiliation{ Thomas Young Centre, University College London, London WC1H 0AH, UK. }
\affiliation{ Interdisciplinary Center for Scientific Computing, University of Heidelberg, Im Neuenheimer Feld 205A, 69120 Heidelberg, Germany }

\author{Angelos Michaelides} 
\email{angelos.michaelides@ucl.ac.uk}
\affiliation{ Thomas Young Centre, University College London, London WC1H 0AH, UK. }
\affiliation{ London Centre for Nanotechnology, Gordon St., London WC1H 0AH, UK. }
\affiliation{ Dept. of Physics and Astronomy, University College London, London WC1E 6BT, UK. }

\author{Dario Alf\`{e}} 
\email{d.alfe@ucl.ac.uk}
\affiliation{ Thomas Young Centre, University College London, London WC1H 0AH, UK. }
\affiliation{ London Centre for Nanotechnology, Gordon St., London WC1H 0AH, UK. }
\affiliation{ Dept. of Earth Sciences, University College London, London WC1E 6BT, UK. }
\affiliation{ Dipartimento di Fisica Ettore Pancini, Universit\`{a} di Napoli Federico II, Monte S. Angelo, I-80126 Napoli, Italy. }


\date{\today}

\begin{abstract}
Fixed node diffusion quantum Monte Carlo (FN-DMC) is an increasingly used computational approach for investigating the electronic structure of molecules, solids, and surfaces with controllable accuracy.
It stands out among equally accurate electronic structure approaches for its favorable cubic scaling with system size, 
which often makes  FN-DMC 
the only computationally affordable high-quality method in large condensed phase systems with more than 100 atoms.
In such systems FN-DMC deploys pseudopotentials to 
substantially improve efficiency.
In order to deal with non-local terms of pseudopotentials, the FN-DMC algorithm must use an additional approximation, leading to the so-called localization error.
However, the two available approximations, the locality approximation (LA) and the T-move approximation (TM), have certain disadvantages and can make DMC calculations difficult to reproduce.
Here we introduce a third approach, called the determinant localization approximation (DLA).
DLA eliminates reproducibility issues and systematically provides good quality results and stable simulations that are slightly more efficient than LA and TM.
When calculating energy differences -- such as interaction and ionization energies -- DLA is also more accurate than the LA and TM approaches. 
We believe that DLA paves the way to the automization of FN-DMC and its much easier application in large systems.
\end{abstract}

\maketitle


\section{Introduction}

A wide range of scientific topics greatly benefit from computer simulations, such as 
crystal polymorph prediction, molecular adsorption on surfaces, 
the assessment of phase diagrams, phase transitions, nucleation, and more.
The accuracy of the computational methods employed for such simulations is of fundamental importance. 
In a wide range of physical-chemical problems 
many  important static, dynamic, and thermodynamic properties are related to the potential energy surface.
Thus, one of the grand challenges of computational modelling is the evaluation of accurate energetics for molecules, surfaces and solids.
%
This challenge is far from straightforward, because 
various types and strengths of
interatomic and intermolecular interactions are relevant, and a method must describe all of them correctly.

There are various computational methods acknowledged for having high accuracy.
For condensed phase systems one very promising methodology is  quantum Monte Carlo (QMC),\cite{foulkes01}
often in the fixed-node (FN) diffusion Monte Carlo (DMC) flavor.
FN-DMC has favorable scaling with system size (between the 3-rd and the 4-th power of the system size) and it can be efficiently deployed on high performance computer facilities.
Nowadays there is an increasing amount of benchmark data for solids and surfaces obtained via FN-DMC.\cite{zen-pnas18, water_graphene_JPCL, Zen_Kao_2016, AlHamdani_hBN_2017, AlHamdani_CNT_2017, AlHamdani_hBN_2015, Trail_TiO2_2017, Luo_TiO2_2017, Dubecky_chemrev2016, Sorella_Cerium_2015, Wagner_hBN_2015, Wagner_VO2_2015, Wagner_cuprates_2014, Azadi_prl2014, Benali:2016ew, Shin:2017kw, Ahn:2018dq, Tsatsoulis_W+LiH_2017, DoblhoffDier_H2+Cu111_2017}
Data provided by DMC is of use in tackling interesting materials science problems and also to  help the improvement of density functional theory (DFT)\cite{TkatchenkoReview2019} and other cheaper computational approaches.

DMC implements a technique to project out the exact ground state wave function $\Phi$ from a trial wave function $\PsiT$ by performing a propagation according to the imaginary time-dependent Schr\"{o}dinger equation.\cite{foulkes01} 
However, an unconstrained projection leads to a bosonic wave function, 
so in fermionic systems the fixed-node (FN) approximation is typically employed to keep the projected wave function antisymmetric. FN-DMC constrains the projected wave-function $\Phi_{FN}$ to have the same nodes as a trial wave function $\PsiT$.\footnote{
Alternative strategies to the FN approximation have been developed,\cite{ReleaseNode_Ceperley1984, AfunDMC_jcp1993, FGFMC_jcp1982, ExactMC_Anderson1991, ExactMC_Zhang1991, ExactMC_Kalos2000, SHDMC_prb2009}
which are typically more accurate but much more demanding in terms of computational cost.
Thus, typically only FN-DMC is affordable in large systems.}
Thus, $\Phi_{FN}$ is as close as possible to the exact (unknown) fermionic ground state $\Phi$ given the nodal constraints, and the equality is reached if $\PsiT$ has exact nodes.
In addition to the FN approximation, in most practical DMC simulations pseudopotentials are used, because the core electrons in atoms significantly increase the computational cost of FN-DMC simulations.\cite{Ceperley_Zscaling_1986, Hammond_Zscaling_1987, Needs_Zscaling_2005} 
There are also a few other technical aspects of FN-DMC that can affect its accuracy and efficiency, such as 
the actual implementation of the imaginary time-dependent Schr\"{o}dinger propagation for a finite time-step.\cite{ZSGMA, umrigar93, depasquale88}
However, in general the most important and sizeable approximations in FN-DMC arise from the fixed-node constraint and the use of pseudopotentials.

The use of pseudopotentials in FN-DMC brings a twofold approximation.
The first and trivial source of approximation is due to the fact that no pseudopotential is perfect. Pseudopotentials (PPs) can represent implicitly the influence of the core electrons only approximatively, and any method employing pseudopotentials will be affected by this issue.
The second source of approximation 
is more subtle and tricky. 
PPs have non-local operators,
i.e., terms $\hat V_{NL}$ such that their application on a generic function $\xi(\R)$ of the electronic coordinates $\R$ gives 
$ \hat V_{NL} \xi(\R) = \int \left< \R \right| \hat V_{NL} \left| \R' \right> \xi(\R') d\R' \,. $
The non-local PP operators are a big issue in FN-DMC simulations. Indeed, one of the quantities that should be evaluated in the FN-DMC projection is the value of the non-local terms applied to the projected wave function, $\hat V_{NL} \Phi_{FN}$, which cannot be calculated as we do not know the functional form of $\Phi_{FN}$.
There have been attempts to circumvent the difficulty 
but the issue persists.\cite{Casula06}
%
There has been no satisfactory way to deal with the non-local PP terms within FN-DMC exactly,
and it is necessary to rely on some approximation. 
So far, there are two alternatives: 
the locality approximation\cite{LA:mitas91} (LA) or the T-move approximation\cite{Casula06, Casula10} (TM).
In the former the trial wave function $\PsiT$ is used to localize the non-local PP terms,
so the unknown term $\Phi_{FN}^{-1} \hat V_{NL} \Phi_{FN}$ is approximated with $\PsiT^{-1} \hat V_{NL} \PsiT$.
In the latter, only the terms in $\hat V_{NL}$ yielding a sign-problem are localized using $\PsiT$.
In both LA and TM there is a {\em localization error}.\footnote{ The localization error goes to zero (ideally) as we can find a $\PsiT$ closer to the (unknown) fixed-node solution $\Phi_{FN}$, both in LA and TM.}
Both LA and TM are used in production calculations and it is unclear which is better.\footnote{\revision{TM yields more stable simulations than LA.\cite{Casula06, Drummond:2016gp} However, if the trial wave function $\PsiT$ is good LA has typically no stability issues. In these cases TM is often affected by larger finite time-step and stochastic errors than LA.\cite{Pozzo:PRB2008} Moreover, LA satisfies a detailed balance condition while TM does not, and the size-consistency condition can be harder to satisfy in TM (indeed, the first version of the TM algorithm\cite{Casula06} was size-inconsistent, an issue solved four years later with two alternative revised TM algorithms\cite{Casula10}).
Recent investigations\cite{Krogel_locerror_2017, Dzubak_locerror_2017} of the localization errors in LA and TM 
do not show that one method is clearly superior to the other in terms of accuracy.}}

%
Both LA and TM have a big problem: reproducibility.
As discussed above, the localization error arises from the projection of all or part of $\hat V_{NL}$ on $\PsiT$.
With either LA or TM, FN-DMC will produce different results with different $\PsiT$, even if the nodes are unchanged.
Unfortunately, $\PsiT$ has a level of arbitrariness in the way it can be defined, because it is not straightforward to tell what is the optimal choice for $\PsiT$. Ideally, we want a $\PsiT$ as close as possible to the exact ground state (which is unknown) and for which the ratio $\PsiT(\R)/\PsiT(\R')$ is quickly evaluated computationally.
In practice, $\PsiT$ can have many different functional forms, 
and different QMC packages often use different forms,  exacerbating the reproducibility problem. 
Even within a specific implementation of $\PsiT$, there are parameters to be set in some way, and  there is no unique way to do this. Moreover, in large systems the number of parameters increases rapidly, leading to additional difficulties.
For this reason, it is not uncommon to find 
differences in FN-DMC results from nominally similar studies.
\revision{
This contrasts with other electronic structure techniques where agreement between different studies is relatively straightforward to achieve. Resolving issues such as this are critical to reducing the labour intense nature of the DMC simulations and making DMC easier to use in general.
}

In this paper we introduce a new approach to deal with the non-local potential terms in FN-DMC which resolves the reproducibility problem.
We call this method the determinant localization approximation (DLA).
In DLA, as explained below, the key point is to do the localization on just the determinant part of the trial wave function.
The localization error in DLA is not eliminated, but it can be reproduced systematically across different implementations of $\PsiT$ and in different QMC packages. 
To this aim, the localization only uses the part of the trial wave function that can be obtained deterministically: the determinant part, which fixes the nodes. 
In this way the reproducibility of DLA is guaranteed by construction.

It needs to be tested if DLA yields results competitive to LA and TM.
It has to be noticed that with PPs we are always interested in energy differences, and not in absolute energies.
So, the most accurate method is not necessarily the one with the smallest absolute localization error, but the method that makes consistently the same localization error across different configurations of the same system, such that there is the largest error cancellation in the energy difference. 
By construction, DLA makes very consistent localization errors.
Indeed, we observe in all the representative cases considered in this paper that DLA always yields accurate results, which are systematically better than LA and TM whenever the $\PsiT$ is not optimal.
Moreover, we notice that DLA produces very stable simulations, in contrast to LA.
In terms of efficiency, DLA appears slightly more efficient than both LA and TM.
All these features make DLA the best candidate to perform FN-DMC calculations in large systems, where the quality of $\PsiT$ could be hard to assess and a stable and efficient simulation is highly needed.

The outline of the paper is the following:
we provide a short overview of the FN-DMC method in section \ref{sec:review-FNDMC}; 
we describe our determinant localization approximation in section \ref{sec:DLA};
we illustrate the results produced by DLA, compared with LA and TM, in section \ref{sec:results}, 
with a specific focus on
interaction energy evaluations (\ref{sec:Etot_Eb}), 
ionization energies (\ref{sec:IE}), stability (\ref{sec:DLAstability}) and efficiency (\ref{sec:DLAefficiency}). 
A reader already familiar with DMC can skip to section~\ref{sec:DLA}.
We draw our final conclusions in section \ref{sec:conclusion}.

\section{Overview on fixed node diffusion Monte Carlo}\label{sec:review-FNDMC}

\subsection{The trial wave function}\label{sec:PsiT}

The trial wave function has a critical role in determining the accuracy of FN-DMC. 
A QMC trial wave function is the product 
$ \PsiTR = \DR * \eJR $
of an antisymmetric function $\DR$ and a symmetric (bosonic) function $\eJR$, called the Jastrow factor, where $\R$ is the electronic configuration.
\revision{
The function $\DR$ is typically a single Slater determinant, especially when large systems are simulated.
However, it is worth mentioning that if the system under consideration is not too large (say, generally not more than a few atoms) better functions can be used, such as multi-determinant expansions of Slater determinants\cite{FilippiUmrigar_JCP1996, Valsson:2010p25419, Zimmerman:2009hh, Toulouse:2008p27527, CIPSI13, Scemama:2018}, valence-bond wave functions\cite{J_valence_bond_wf_JCP2011}, the antisymmetrized geminal product\cite{JAGP_Casula2003, JAGP_diradicals_Zen2014}, the Pfaffian\cite{Pfaffian_prl2006}, and others (see for instance the review by \citet{Lester_ChemRev2012}).
Moreover, the backflow transformation\cite{Backflow_Feynman_1956, InhomogeneousBackflow_PRE2006, Orbital_dependent_backflow_PRB2019} can be employed to further improve any of the aforementioned ans\"{a}tze, at the price of a significantly larger computational cost.
}
The Jastrow factor describes the dynamical correlation between the electrons, 
by including explicit functions of the electron-electron distances.
%
\revision{
In DMC a property of interest is the nodal surface of $\PsiT$, which is the hypersurface corresponding to $\PsiTR=0$, for real wave functions, or the complex phase of $\PsiT$ for complex wave functions.
They are both determined by $\DR$, as the Jastrow factor can only alter the amplitude of $\PsiT$.
}
The Jastrow factor $\J$ is implemented differently in different QMC  packages.

When large and complex systems are simulated, such as adsorption on surfaces or molecular crystals, the most common practice is to obtain $\D$ from a deterministic approach, usually DFT, and to decide a functional form for $\J$ and optimize, within the variational Monte Carlo (VMC) scheme,\cite{foulkes01} the parameters minimizing either the energy or the variance.
Since $\D$ comes from a deterministic method, there is no reproducibility problem here, and in taking energy differences we can usually expect a large cancellation of the FN error.
On the other hand, $\J$ is optimized stochastically, so its parameters are affected by an optimization uncertainty. 
Dealing with this uncertainty becomes increasingly challenging as the system gets larger.
Moreover, a new optimization of $\J$ is needed for every distinct orientation of the molecular systems, and optimizing $\J$ so frequently is tedious, timeconsuming, and, due to the stochastic nature of the optimisation procedure, can lead to Jastrow factors of different qualities, resulting in less than optimal cancellation of errors.
\revision{
A human supervision of the optimization is always highly recommended, if not necessary.
The optimization is responsible for making QMC labour intense and non automatic. 
}

\subsection{Diffusion Monte Carlo}


The DMC algorithm with importance sampling performs a time evolution of $\fRt=\PsiTR\psi(\R,t)$,
where $\PsiTR$ is a trial wave function (described in Section \ref{sec:PsiT}), $\R$ are the $3N$-dimensional electronic coordinates and $\psi(\R,t)$ is the solution at time $t$ of the imaginary time Sch\"{o}dinger equation
\begin{equation}\label{eq:itimeSchoringer}
-{\partial \over \partial t} \psi(\R,t) = \left( \hat H - E_T \right) \psi(\R,t) 
\,,
\end{equation}
where $\hat H$ is the Hamiltonian and $E_T$ a trial energy, 
with initial condition $\psi(\R,0)=\PsiTR$ and converging exponentially to the exact ground state $\Phi(\R)$ for $t\to\infty$.
Thus, $\lim_{t\to\infty} f(\R,t)=\PsiTR\Phi(\R)$.
Since $\Phi$ is an eigenstate for $\hat H$, the ground state energy $E_0$ can be calculated using the mixed estimator:
\begin{equation}\label{eq:e0_mix}
E_0 =
{\left< \Phi \right| \hat H \left| \Psi_T \right> \over \left< \Phi \middle| \Psi_T \right>} = 
{ \int \Phi(\R) \PsiTR E_L(\R) d\R \over \int \Phi(\R) \PsiTR d\R } 
\, ,
\end{equation}
where 
$
E_L(\R) 
= {\left< \R \right| \hat H \left| \Psi_T \right> \over \left< \R \middle| \Psi_T \right>} 
$
is the local energy in the electronic configuration $\R$ for the trial wave function $\Psi_T$.

The time evolution of $\fRt$ follows from the imaginary time Sch\"{o}dinger equation (\ref{eq:itimeSchoringer}), 
which in integral form leads to:
\begin{equation}\label{eq:fRtevolution}
f({\bf R}',t+\tau) =\int G(\R' \leftarrow \R,\tau) \fRt d\R
\end{equation}
where $\tau$ is the time-step, $G(\R' \leftarrow \R,\tau)$ is the Green function for the importance sampling, which is defined (symbolically) as:
\begin{equation}\label{eq:Gdef}
G(\R' \leftarrow \R,\tau) \equiv 
{ \PsiTRp \over \PsiTR } \left< \R' \right| \exp(-\tau \hat H) \left| \R \right> \,.
\end{equation}
Thus, by starting from $f({\bf R},0)=\PsiTR^2$ and performing an evolution according to the Green function $G(\R' \leftarrow \R,t)$ 
we are able to assess expectation values of the exact ground state $\Phi$:
\begin{equation}\label{eq:DMCprojection}
\Phi({\bf R'})\PsiTRp = \lim_{t\to\infty} \int G(\R' \leftarrow \R,t) \PsiTR^2 d\R
\, .
\end{equation}
This is the process implemented in the DMC algorithm. 
In fermionic systems the fixed-node (FN) approximation is typically introduced, 
so the FN Hamiltonian 
$\hat H_{FN} \equiv \hat H + \hat V_{FN}$, 
where $\hat V_{FN}$ is an infinite wall at the nodal surface of $\Psi_T$, is used.
Further details are reported in Appendix~\ref{app:DMCimplementation}.

The Hamiltonian $\hat H$ is the sum of the kinetic and potential operators $\hat K$ and $\hat V$, respectively.
In all-electron calculations the potential operator $\hat V$ is local, $\hat V=\hat V_L$.
However, in general there is the need to deploy pseudopotentials to represent the core electrons of the atoms and reduce the computational cost of the calculation, see Appendix~\ref{app:AEvsPP}.
In this case the potential term has both local and non-local operators: $\hat V=\hat V_L+\hat V_{NL}$.
%
The presence of non-local operators in the potential complicates the formulation of the DMC algorithm and forces the introduction of a further approximation.
In the following we will first 
consider the simple case of a potential with only local operators, sec.~\ref{sec:Vloc},
and later we will consider the case of potential term with non-local operators, sec.~\ref{sec:Vnonloc}.

\subsection{Green's function for $\hat H = \hat K + \hat V_L$}\label{sec:Vloc}


The simplest case is when the Hamiltonian has only a local potential term, thus it can be written as
$\hat H = \hat K + \hat V_L$, 
with $\hat K = -{1\over 2} \nabla^2$.
By substitution in the imaginary time Schr\"{o}dinger equation (eq. (\ref{eq:itimeSchoringer})),
multiplication 
by $\PsiTR$, and some algebraic operations,
we obtain:
\begin{equation}\label{eq:itimeFRt_loc}
{\partial \over \partial t} \fRt = {1\over 2} \nabla^2 \fRt - \nabla \cdot \left( \VR \fRt \right) 
- \left[ E_L(\R) - E_T \right] \fRt  
\,,
\end{equation}
where $\VR = \nabla \log \left| \PsiTR \right|$.
Thus, the time evolution of $\fRt$ is given on the right hand side (RHS) of eq. (\ref{eq:itimeFRt_loc}).
If the RHS only had the first two terms, we would have a pure drift-diffusion process, 
having a Green's function that for a small time-step $\tau$, and for $N$ electrons in the system, can be approximated as:
$
G_{DD}(\R' \leftarrow \R,\tau) = { 1 \over (2 \pi \tau)^{3 N/2}} \exp\left\{ - {  [ \R' - \R - \tau \VR ]^2 \over 2\tau } \right\} \,.
$
The last term on the RHS of eq.~(\ref{eq:itimeFRt_loc}) is the branching term, and its associated Green's function is:
$
G_{B}(\R' \leftarrow \R,\tau) = \exp\left\{ - \tau {  E_L(\R') + E_L(\R) - 2 E_T \over 2 } \right\} \,.
$
The Green's function of $\fRt$ for a small time interval $\tau$ can be approximated\cite{book_QMC_Hammond_Lester_Reynolds, Book_MonteCarlo_Kalos_Whitlock_2009} as:
\begin{equation}\label{eq:G_BDD}
G_{BDD}(\R' \leftarrow \R,\tau) \approx G_{B}(\R' \leftarrow \R,\tau) G_{DD}(\R' \leftarrow \R,\tau) \,,
\end{equation}
which is exact for $\tau\to0$.
$G_{BDD}(\R' \leftarrow \R,\tau)$ can be used to approximate the Green's function for an arbitrarily large time interval $t$.
\footnote{
Given the values of $t$ and $\tau$, we can approximate  
$
G(\R' \leftarrow \R,t) \approx 
\int \prod_{i=1}^{n} G(\R_i \leftarrow \R_{i-1},\tau) d{\R_{n-1}} \ldots d{\R_{1}}
\, ,
$
where $n=t/\tau$, $\R'=\R_n$, and $\R=\R_0$.
}
$G_{BDD}$ defines a branching-drift-diffusion process, as described for instance in Ref.~\onlinecite{foulkes01}.
The algorithms implemented in QMC packages are usually a little more involved.\footnote{ 
For instance, it has been observed that the time-step error is largely decreased if a Metropolis step is included in order to enforce detailed balance\cite{CeperleyKalosLebowitz:1981, ReynoldsEtAl:1982}, and that it is convenient to reformulate the algorithm in order to implement electron-by-electron updates instead of configuration-by-configuration updates, as it improves the efficiency in large systems\cite{CeperleyKalosLebowitz:1981, ReynoldsEtAl:1982}.
Moreover, in the proximity of the nodes of $\PsiTR$ the local energy $E_L(\R)$ and the drift vector $\VR$ diverge, yielding instabilities in the branching and drifting terms. This issue is substantially reduced by considering modified versions of the branching and drifting terms close to the nodes\cite{umrigar93, ZSGMA}.
The Metropolis step,  the modification to the branching and drifting terms, and the electron-by-electron DMC algorithm do affect the simulations for finite value of $\tau$, but do not change the limit for $\tau\to 0$.
They are very important technical aspects, as they enhance the stability and efficiency of the algorithm. For instance the development of a size consistent version of the modification to the branching term, as discussed in Ref.~\onlinecite{ZSGMA}, yield a speedup of up to two orders of magnitude \revision{in the evaluations of binding and cohesive energies\cite{ZSGMA, zen-pnas18}} 
}
However, there is no need here to complicate further the picture. 
We will be concerned with the results of DMC in the continuous limit $\tau\to 0$. 
In this limit, the only bias in the DMC energy evaluation $E_{FN}$
is given by the FN approximation.
In particular, $E_{FN} \ge E_0$, with the equality reached if the nodes of $\Psi_T$ are exact.

\subsection{Green's function for $\hat H = \hat K + \hat V_L + \hat V_{NL}$}\label{sec:Vnonloc}


When pseudopotentials are used the potential term has non-local operators $\hat V_{NL}$, and 
the Hamiltonian can be written as  
$\hat H = \hat K + \hat V_L + \hat V_{NL}$.
If we consider the imaginary time Schr\"{o}dinger equation (\ref{eq:itimeSchoringer}) and substitute $\hat H$,
we obtain the following time evolution of $\fRt$:
\begin{equation}\label{eq:itimeFRt}
{\partial \over \partial t} \fRt = {1\over 2} \nabla^2 \fRt - \nabla \cdot \left( \VR \fRt \right) 
- \left[ { \left( \hat K + \hat V_L \right) \PsiTR \over \PsiTR } 
	+ { \hat V_{NL} \psi(\R,t) \over \psi(\R,t) } 
	- E_T 
\right] \fRt  \,.
\end{equation}
The drift and diffusion terms on the RHS are identical to eq. (\ref{eq:itimeFRt_loc}),
but there is a complication in the branching term.
Indeed,  we cannot calculate $\hat V_{NL} \psi(\R,t) \over \psi(\R,t)$,
as we do not know the analytical form of $\psi(\R,t)$.

There is an alternative approach, which is to write the Green's function $G(\R' \leftarrow \R,\tau)$ for $\hat H$.
Using the Zassenhaus formula, for small $\tau$ we can approximate 
$e^{-\tau (\hat K + \hat V_L + \hat V_{NL})}$ 
with
$e^{ -\tau \hat V_{NL}} e^{-\tau (\hat K + \hat V_L)}$,
and by substituting it into eq. (\ref{eq:Gdef}) 
we obtain 
\begin{equation}\label{eq:G_NL}
G(\R' \leftarrow \R,\tau) \sim 
\int T_{NL}(\R' \leftarrow \tilde \R,\tau) * G_L(\tilde \R \leftarrow \R,\tau) d\tilde\R
\end{equation}
where  
$
G_L(\R' \leftarrow \R,\tau) \equiv 
{ \PsiTRp \over \PsiTR } \left< \R' \right| e^{-\tau (\hat K + \hat V_L)} \left| \R \right> 
$
is the Green's function for the local part of the Hamiltonian, which has been discussed in the previous section, 
and 
$
T_{NL}(\R' \leftarrow \R,\tau) \equiv 
{ \PsiTRp \over \PsiTR } \left< \R' \right| e^{-\tau \hat V_{NL}} \left| \R \right> 
$
is the Green's function of the non-local part of the potential.
For small $\tau$ we have that
$ 
T_{NL}(\R' \leftarrow \R,\tau) \sim \delta_{\R',\R} - \tau V_{\R',\R}
$
where $ \delta_{\R',\R}$ is the Dirac's delta and 
\begin{equation}\label{eq:Vrr'}
V_{\R',\R} \equiv { \PsiTRp \over \PsiTR } \left< \R' \right| \hat V_{NL} \left| \R \right> \,.
\end{equation}
Notice that $V_{\R',\R}$ can be either positive or negative depending on $\PsiT$, $\hat V_{NL}$, $\R$ and $\R'$.
Whenever $V_{\R',\R}>0$ for some $\R'\ne\R$, then 
$T_{NL}(\R' \leftarrow \R,\tau)<0$.
The DMC algorithm needs to interpret the Green's function as a transition probability, 
but if $T_{NL}(\R' \leftarrow \R,\tau)<0$ for some $\R$ and $\R'$, 
it cannot be a transition probability from $\R$ to $\R'$ (sign problem).
%
Thus, the presence of $\hat V_{NL}$ yields a sign problem in the DMC algorithm\cite{Casula06, Casula10},
because it gives a Green's function $G(\R' \leftarrow \R,\tau)$ which can have negative terms.

There is no direct solution to this problem, and 
as a consequence an approximation is introduced.
As noted earlier two approaches are available:
either to use the locality approximation (LA) \cite{LA:mitas91}
or Casula's T-move approximation (TM) \cite{Casula10, Casula06}.
They are summarized in the following two sections.

\subsubsection{Locality approximation in FN-DMC}\label{sec:LA}

The approach taken in LA is to approximate the unknown quantity 
${\hat V_{NL} \Phi(\R,t) \over  \Phi(\R,t)}$
with  
${\hat V_{NL} \Psi_T(\R) \over  \Psi_T(\R)}$,
which is the value of the non-local potential localized on the trial wave function $\PsiTR$.
By using this approximation in eq. (\ref{eq:itimeFRt}) we obtain that the 3rd term on the RHS is
$
-[E_L(\R) - E_T] f(\R,t) \,,
$
and the equation becomes identical to eq. (\ref{eq:itimeFRt_loc}).
Thus, the Green's function in LA is given by eq. (\ref{eq:G_BDD}) and the DMC algorithm is a branching-drift-diffusion process.

The major difference from section \ref{sec:Vloc}
is that we approximate the Hamiltonian, 
which is no longer given by the FN Hamiltonian $\hat H_{FN} \equiv \hat H + \hat V_{FN}$, 
but by 
\begin{equation}\label{eq:H-LA}
\hat H^{LA}_{FN} \equiv \hat K + \hat V_L + {\hat V_{NL} \Psi_T \over \Psi_T} + \hat V_{FN} \,,
\end{equation}
where the notation ${\hat V_{NL} \Psi_T \over \Psi_T}$
is used to indicate that the non-local potential $\hat V_{NL}$ has been localized using the function $\PsiTR$. So, given a generic function $\xi(\R)$, we have
$
{\hat V_{NL} \Psi_T \over \Psi_T} \xi(\R) = 
\int d\R' \Psi_T(\R') \left< \R' \right| \hat V_{NL} \left| \R \right> 
{\xi(\R) \over \Psi_T(\R)}
\,.
$
Notice that $\hat H^{LA}_{FN}$ has no non-local potential term, 
i.e. the action of $\hat H^{LA}_{FN}$ on the generic function $\xi$ at point $\R$ only depends on the value of $\xi$ at $\R$. 

The ground state for $\hat H^{LA}_{FN}$ is the projected wave function $\Phi_{FN}^{LA}$.
The expectation value of the energy $E_{FN}^{LA}$ can be evaluated using the mixed estimator,
because ${\hat V_{NL} \Psi_T \over \Psi_T} \Psi_T(\R) = \hat V_{NL} \Psi_T(\R)$,
so $\hat H^{LA}_{FN} \left| \Psi_T \right> = \hat H \left| \Psi_T \right>$.
However, in general $\Phi_{FN}^{LA}$ is different from the (unknown) ground state $\Phi_{FN}$ for $\hat H_{FN}$, thus $E_{FN}^{LA} \ne E_{FN}$.
In other words, with LA we have lost the variationality of the approach,
because the error introduced by this approximation can either be positive or negative,
and $E_{FN}^{LA}$ is not, in general, an upper bound for $E_0$.
Only in the (ideal) case of  $\Psi_T = \Phi_{FN}$ we do have 
$\hat H_{FN}^{LA}  \left| \Phi_{FN} \right> = E_{FN} \left| \Phi_{FN} \right>$,
so $E_{FN}^{LA} = E_{FN}$.
As a corollary, with the exact trial wave function, 
$\Psi_T = \Phi$,
then we have that 
$E_{FN}^{LA} = E_0$.
However, the trial wave function having exact nodes is not a sufficient condition for having $E_{FN}^{LA} = E_0$,
as the LA depends on the overall trial wave function $\Psi_T$, and not just on its nodes.\footnote{
If $\Psi_{T1}$ and $\Psi_{T2}$ are two different trial wave functions with the same nodes (e.g., having the same determinant $\DR$ but different Jastrow factors), the corresponding FN-DMC-LA energies, 
$E_{FN}^{LA}[\Psi_{T1}]$ and $E_{FN}^{LA}[\Psi_{T2}]$, will be in general different,
$E_{FN}^{LA}[\Psi_{T1}] \ne E_{FN}^{LA}[\Psi_{T2}]$,
because of the different localization errors.
}
In other words, $E_{FN}^{LA}$ has both a FN error and a localization error.
\revision{
\citet{LA:mitas91} showed that the error (including both FN and localization) on the energy evaluation is quadratic in the wave function error, i.e., $ E_{FN}^{LA} - E_0 = {\cal O}[(\PsiT - \Phi)^2]$.
}
Notice that FN and localization errors add to the time-step error and are present also in the limit of zero time-step.

\subsubsection{T-move approximation in FN-DMC}\label{sec:TM}

In the T-move approach, the non-local Green's function $T_{NL}$ includes only the terms without sign-problems and the remaining part of the non-local potential is localized in a similar manner to LA.
To this aim, the positive, $V_{\R',\R}^+$, and negative, $V_{\R',\R}^-$ parts, 
$ V_{\R',\R}^{\pm} = {V_{\R',\R} \pm |V_{\R',\R}| \over 2} $,
of the term $V_{\R',\R}$ defined in eq. (\ref{eq:Vrr'}) are used.
The sign problem arises from terms $V_{\R',\R}^+$, which have to be localized, while the terms $V_{\R',\R}^-$ yield a $T_{NL}$ with non-negative sign. 
Details about how this algorithm can be implemented are discussed in Refs.~\onlinecite{Casula06, Casula10}. 

The corresponding T-move Hamiltonian is:
\begin{equation}\label{eq:H-TM}
\hat H^{TM}_{FN} \equiv \hat K + \hat V_L + \hat V_{NL}^- + {\hat V_{NL}^+ \Psi_T \over \Psi_T} + \hat V_{FN} \,,
\end{equation}
where the operators $\hat V_{NL}^+$ and $\hat V_{NL}^-$ correspond to 
$V_{\R',\R}^+$ and $V_{\R',\R}^-$, respectively.
The projected wave function $\Psi_{FN}^{TM}$ is the ground state of $\hat H^{TM}_{FN}$, and since 
$\hat H^{TM}_{FN} \ket{\Psi_T} = \hat H \ket{\Psi_T}$
the expectation value of the energy $E^{TM}_{FN}$ can be evaluated using the mixed estimator.
Similar to the LA approach, the projected function $\Phi_{FN}^{TM}$ is in general different from the fixed node ground state $\Phi_{FN}$,
but if the trial wave function $\Psi_T=\Phi_{FN}$ then $E^{TM}_{FN}=E_{FN}$.
If 
$\Psi_T=\Phi$
then $E^{TM}_{FN}=E_0$, 
but if $\PsiT$ has exact nodes but differs from $\Phi$ then $E^{TM}_{FN} \ne E_0$.
Similar to LA, TM depends on the overall trial wave function $\Psi_T$.\footnote{ 
Two different trial wave functions  $\Psi_{T1}$ and $\Psi_{T2}$ with the same nodes have in general different FN-DMC-TM energies 
$E_{FN}^{TM}[\Psi_{T1}] \ne E_{FN}^{TM}[\Psi_{T2}]$,
because their localization errors are different.
}
Note that the FN and localization errors add to the time-step error, and are present also in the limit of zero time-step.

The TM approach is computationally slightly more expensive than LA and often has a larger time-step error.
However, it has two advantages over LA:
$E^{TM}_{FN}$ is an upper bound of the exact ground state $E_0$,\cite{tenHaaf:1995} 
and it is a more stable algorithm than the LA.

\section{New approach: determinant localization approximation in FN-DMC: }\label{sec:DLA}


The major practical disadvantage of both LA and TM is that the results are highly dependent on the Jastrow factor $\J$.
This might result in problems of reproducibility, especially between results from different QMC packages, as the Jastrow factor is often expressed in different and non equivalent functional forms across the codes.
Moreover, 
the parameters of the Jastrow are affected by stochastic uncertainty. 
In contrast, it is much easier to control the reproducibility of the determinant part of the wave function $\D$, which is generally obtained from a deterministic method, 
e.g. DFT.

Therefore, we propose to use only the determinant part $\D$ of the trial wave function to localize the non-local potential $\hat V_{NL}$.\footnote{\revision{Notice that this approach was used three decades ago by \citet{Hammond_Zscaling_1987} in one of the first works using PPs in FN-DMC, albeit the motivation was to make runs cheaper. Shortly after the LA was introduced\cite{LA:mitas91} -- with a prescription to solve the integrals involved in the localization (see Appendix~\ref{app:JvsPP} and eq. \ref{eq:Vncloc} therein) -- and it quickly became the default approach.}}
If we bear in mind that pseudopotentials are tested by developers using deterministic methods -- density functional theory\cite{BFD, trail05_SRHF} or coupled cluster with single, double and perturbative triple excitations (CCSD(T))\cite{CEPP, CEPP-3d, eCEPP, ccECP} --
our suggestion seems also quite reasonable, because they are not tested widely and systematically in the presence of a Jastrow and within a DMC scheme.
 
In DLA the FN Hamiltonian is:
\begin{equation}\label{eq:H-DLA}
\hat H^{DLA}_{FN} \equiv \hat K + \hat V_L + {\hat V_{NL} \D \over \D} + \hat V_{FN} \,,
\end{equation}
and the associated projected wave function is $\Phi_{FN}^{DLA}$.
In order to be able to use the mixed estimator, we need to define the Hamiltonian:
\begin{equation}\label{eq:H-DLAval}
\hat H^{DLA} \equiv \hat K + \hat V_L + {\hat V_{NL} \D \over \D} \,,
\end{equation}
such that 
$\hat H^{DLA}_{FN} \ket{\PsiT} = \hat H^{DLA} \ket{\PsiT}$
and
\begin{equation}\label{eq:eFN-DLA_mix}
E_{FN}^{DLA} =
{ \int \Phi_{FN}^{DLA}(\R) \PsiTR E_L^{DLA}(\R) d\R \over \int \Phi_{FN}^{DLA}(\R) \PsiTR d\R } 
\,,
\end{equation}
where the local energy is
$E_L^{DLA}(\R) = { \bra{\R} \hat  H^{DLA}_{FN} \ket{\PsiT} \over \braket{\R}{\PsiT} }$.
\revision{
DLA becomes exact in the limit of $\D\to\Phi$, as 
$H^{DLA}_{FN} \ket{\Phi} = \hat H \ket{\Phi}$ and 
$E^{DLA}_{FN} = E_0$.
It can be shown, along the lines of the argument of \citet{LA:mitas91}, that the error $E_{FN}^{DLA} - E_0$ on the energy evaluation (including both FN and localization) is ${\cal O}[(\D - \Phi)^2]$.
Comparing with the corresponding error within LA, it implies than the DLA error on the absolute energy is typically expected to be larger than the LA error, as $\PsiT=\D\eJ$ is typically closer than $\D$ to $\Phi$.
On the other hand, we suggest and alternative point of view: the DLA approach should be seen as a modification of the PPs. PPs introduce an approximation in the Hamiltonian, whereas the remaining part of $\hat H$ comes from first principles (see Appendix~\ref{app:AEvsPP}). 
There is no proof showing that PPs provide a better approximation of the core if the localization is performed using a wave function with or without the Jastrow factor. However, without  the Jastrow there are clear advantages in terms of reproducibility of results and better error cancellation in energy differences, as discussed below.
In other words, we are not concerned that $E^{DLA}_{FN}$ might be further from $E_0$ than $E^{LA}_{FN}$ (or $E^{TM}_{FN}$) as long as $\hat V_L + {\hat V_{NL} \D \over \D}$ yields a good representation of the ionic potential energy.
}

Within DLA, the quality of the fixed node energy $E_{FN}^{DLA}$ depends exclusively on $\D$.
The Jastrow factor $\J$ does not affect the accuracy; 
the only influence of the Jastrow is on the efficiency, as it will affect the time-step errors and the variance of $H^{DLA}$.
In the limit of zero time-step all calculations which use the same $\D$ will provide the same energy, no matter what (if any) Jastrow factor is used.\footnote{
If one is concerned with the variationality of the approach, it is also possible to use TM in conjunction with DLA. 
The corresponding Hamiltonian is then:
$
\hat H^{DLTM}_{FN} \equiv \hat K + \hat V_L + \hat V_{NL}^{-} + {\hat V_{NL}^{+} \D \over \D} + \hat V_{FN} \,.
$
Within this second approach we have that $E_{FN}^{DLTM}\ge E_0$, with the equality obtained for $\D=\Phi$.
}

\revision{ 
The implementation of DLA is straightforward, as it is a simplification of the LA algorithm.
It implies the numerical integration of $\D$ (instead of $\PsiT$ in LA) over a sphere to determine the nonlocal potential energy ${\hat V_{NL} \D \over \D}$, see eq.~\ref{eq:Vncloc} in Appendix~\ref{app:JvsPP}. The numerical integration scheme employs quadrature rules, hence a number of wave function ratios on the integration grids are evaluated at every energy measurement.\cite{LA:mitas91}
While in the calculations reported in this manuscript we used this simple implementation, 
it should be noticed that DLA allows a more involved but much more efficient implementation.
Whenever $\D$ is used instead of $\PsiT$, the integrals in eq.~\ref{eq:Vncloc} can be factorized into simpler integrals involving the molecular orbitals defining $\D$.\cite{Hammond_Zscaling_1987}
Thus, all the non-local integrals can be done analytically (e.g., analytical expressions have been obtained by \citet{Hammond_Zscaling_1987} under the assumption that the basis functions are gaussian type orbitals) or numerically, becoming a local potential which, for instance, can be precomputed on a grid at the beginning of a QMC simulation.
This approach prevents the evaluation of many wave function ratios and possibly yields an appreciable speedup.
}

\section{Results}\label{sec:results}

In the previous sections we have outlined  that both LA and TM yield total energies, $E_{FN}^{LA}$ and $E_{FN}^{TM}$, affected by the quality of the trial wave function $\PsiT=\D*\eJ$.
Within the DLA scheme introduced here the total energy $E_{FN}^{DLA}$ is affected only by the determinant part $\D$ of the trial wave function $\PsiT$.
Therefore, DLA eliminates the uncertainty due to the Jastrow factor on the DMC results performed with pseudopotentials.
We are going to show here, in a few examples, the amount of uncertainty that the Jastrow can introduce in LA and TM, in contrast to DLA which is not affected by this uncertainty.

\subsection{DLA is good for interaction energy evaluations}\label{sec:Etot_Eb}

\begin{figure*}[htb]
\centerline{
\includegraphics[width=6.4in]{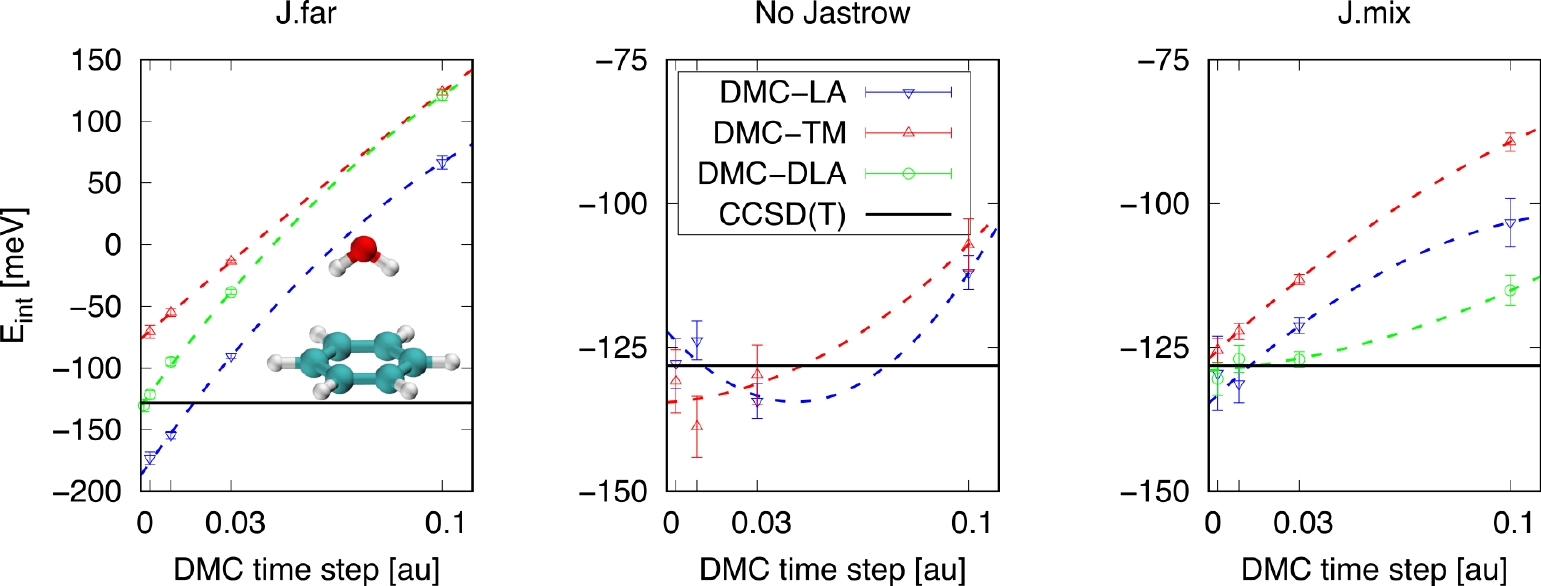}
}
\caption{\label{fig:Eb_W-Bz}
Interaction energy $E_\text{int}$ for a water-benzene complex in the so called ``2-leg'' configuration, for the geometry shown in the inset.
The plots show FN-DMC evaluations versus the time-step $\tau$, using LA, TM, and DLA.
(Left plot) Results obtained using the Jastrow factor J.far in both the bound and the far configurations. J.far has been optimized for the far configuration, i.e. for non-interacting water-benzene molecules.
(Middle plot) Results obtained using a trial wave function without any Jastrow factor. In this case LA and DLA are equivalent.
(Right plot) Results obtained with mixed Jastrow factor: for the non-interacting configuration we used J.far, for the bound configuration we used J.bound.
The reference CCSD(T) value is $-128\pm 1$~meV.\cite{water_graphene_JPCL}.
}
\end{figure*}

The first system that we considered is water bound to benzene, as shown on the inset of Figure~\ref{fig:Eb_W-Bz}.
This is a simple example of the calculation of an interaction energy $E_\text{int} \equiv E_\textrm{bound} - E_\textrm{far}$, which is the difference between the energy ($E_\textrm{bound}$) of the system in the bound configuration and the energy ($E_\textrm{far}$) of the molecules far away. 
Many of these calculations are performed to evaluate a binding energy curve, which are needed for example in adsorption energy calculations of molecules on surfaces.\cite{Zen_Kao_2016, water_graphene_JPCL, AlHamdani_hBN_2017, AlHamdani_CNT_2017, Tsatsoulis_W+LiH_2017, DoblhoffDier_H2+Cu111_2017}
Whereas in this small system it is not overly burdensome to optimize $\J$ at every different geometry, 
in a larger and more complex adsorption system this would be tedious and time-consuming. 
Notwithstanding the variability of the quality of the optimization, due to its stochastic nature. 

\begin{figure*}[htb]
\centerline{
\includegraphics[width=6.4in]{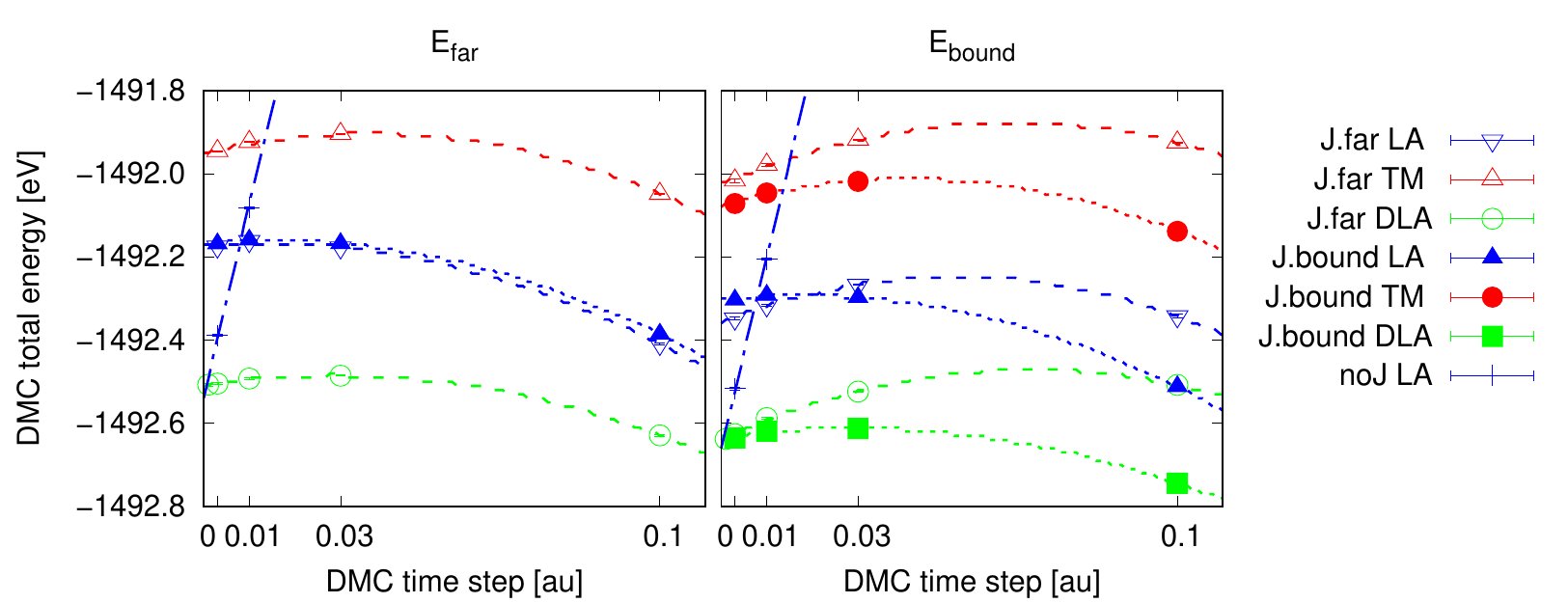}
}
\caption{\label{fig:Etot_W-Bz} \revision{
Total energy for the far ($E_\text{far}$, left plot) and bound ($E_\text{bound}$, right plot) configurations.
The plots show FN-DMC evaluations versus the time-step $\tau$, using LA, TM, and DLA, and different choices of the Jastrow factor: J.far, J.bound, and no Jastrow (noJ), as defined in the text.
FN-DMC-TM noJ points fall outside the interval range ($E_\text{far}=-1491.225(7)$~meV and $E_\text{bound}=-1491.358(1)$~meV in the limit $\tau\to 0$).
}}
\end{figure*}

This specific water-benzene configuration has a reference $E_\textrm{int}$ of -128$\pm$1~meV,\cite{water_graphene_JPCL} as obtained from basis set converged CCSD(T) calculations.\footnote{\revision{
The reference value of -128$\pm$1~meV was obtained  in Ref.~\onlinecite{water_graphene_JPCL} using all-electrons. 
We performed here a new CCSD(T) calculation using the TN-DF pseudopotentials\cite{trail05_SRHF} -- the same used for the DMC evaluations -- and we obtained an interaction of -130$\pm$4~meV. The agreement between the two evaluations indicates that differences between CCSD(T) and DMC should not come from the quality of the employed PPs.}}
A standard setup for FN-DMC was used, with TN-DF pseudopotentials,\cite{trail05_SRHF} a Slater-Jastrow $\PsiT$ with determinant $\D$ obtained from a DFT calculation.\footnote{ 
DFT calculations were performed using the PWSCF package\cite{espresso} with plane wave cutoff of 600~Ry and the LDA functional. Molecular orbitals were later converted into splines\cite{splines-alfe04} to enhance the efficiency in QMC calculations. 
QMC calculations were performed using the CASINO package.
}
A Jastrow factor $\J$ having explicit electron-electron (e-e), electron-nucleus (e-n) and electron-electron-nucleus (e-e-n) terms was used here.
Within this specific functional form of $\J$ we obtained two different Jastrow factors, that we call J.bound and J.far.
The former, J.bound, is the Jastrow obtained when we optimize the parameters by minimizing the variational variance $\Var[E_L]_\text{VMC}$ of the local energy for the bound configuration. 
The latter, J.far, is instead obtained by optimizing the parameters on a configuration where the water and the benzene are far away (around 10~\AA) and are effectively non interacting.
In Figure~\ref{fig:Eb_W-Bz} we compare the reference value with the FN-DMC evaluations obtained with LA, TM, and DLA, and the different Jastrow factors\revision{, whereas Figure~\ref{fig:Etot_W-Bz} reports the FN-DMC total energies.}

%
The left panel of Figure~\ref{fig:Eb_W-Bz} shows results for J.far used for both the bound and the far-away configuration.
With this setup DLA is the only method that provides a reliable interaction energy, which we can estimate to be 
$-131\pm2$~meV for the $\tau\to0$ limit from a quadratic fit of the values obtained at finite values of $\tau$.\footnote{The error bar is the error of the fit, which is probably underestimated \revision{as we are fitting a quadratic function (so, having three parameters) with four points. As a comparison, the actual DLA evaluation with smallest $\tau$, which is $0.001$~au, has a stochastic error due to DMC sampling of $\pm5$~meV.}} 
The estimated $\tau\to0$ limit for LA and TM are 
$-187\pm 3$~meV and 
$-76\pm 1$~meV.\footnote{The reported error comes from the fit.}
So, with this non-optimal Jastrow factor
LA severely over-binds and TM under-binds.\footnote{\revision{Notice that in both LA and TM the bias is dominated by the error in the total energy of the bound system, since the wave function is by construction much better for the unbound systems. Thus, the sign of the TM error can be rationalized: TM under-binds because of the variationality of the approach, which leads to an underestimate of the total energy of the bound system when a suboptimal Jastrow is used. In LA the optimization bias could in principle have any sign, because the approach is not variational. In this specific case the worse Jastrow leads to a lower total energy (see Figure~\ref{fig:Etot_W-Bz}).}}

A different choice, which is indeed the standard procedure adopted in DMC, is to optimize the Jastrow factor specifically for each configuration, i.e. we use J.bound for the bound configuration and J.far for the far configuration. We named this scheme J.mix, and the results obtained with LA, TM, and DLA  are shown on the right panel of Figure~\ref{fig:Eb_W-Bz}.
In this case all three methods are in decent agreement with the CCSD(T) reference, from a quadratic fit we obtain the $\tau\to0$ limit:
$-135\pm 3$~meV for LA,
$-127\pm 1$~meV for TM, and
$-129\pm 2$~meV for DLA.\footnote{
The reported errors are from the fit, and are probably underestimated \revision{because we are fitting three parameters with four points}, especially for the LA approach, were the simulation with the smallest time-step, $\tau=0.003$~au, was very unstable and we could not make the stochastic error smaller than 5~meV. }
The figure also shows the time-step error associated with the three different methods.
The first consideration is that the better choice of the Jastrow has greatly improved the accuracy for any finite $\tau$ evaluation with respect to the case with J.far.
The best time-step dependence is obtained for the DLA approach, where the interaction energy evaluation for $\tau=0.03$~au is $-127\pm1$~meV, which appears already converged (2~meV difference with respect to the $\tau\to0$ limit).

%
At this point one could wonder what is the outcome if the Jastrow is not used at all.
We performed the FN-DMC simulation with LA and no Jastrow (so, it is equivalent to DLA), and the outcome is reported on the middle panel of Figure~\ref{fig:Eb_W-Bz}. 
The $\tau\to0$ limit of the interaction energy is $-125\pm4$~meV, in excellent agreement with the other DMC-DLA evaluations with J.far and J.mix (as it has to be by construction), and also with the reference CCSD(T).
Quite unexpectedly, we also notice that the time-step error is quite small, much smaller than the case with J.far, and similar to the case with J.mix.
This happens despite the huge time-step error on the total energy evaluations \revision{(see Figure~\ref{fig:Etot_W-Bz})} when a Jastrow is not used,
and indicates an unexpectedly good error cancellation of the finite step bias in the energy difference.
We do not know if this behavior of the no Jastrow case is transferable to other systems.
If it was, one would be tempted to do simulations without a Jastrow.
However, this is not recommended because the variance of the local energy $\Var[E_L]$
is much larger without a Jastrow, around ten times the variance of the Slater-Jastrow $\PsiT$.
Since the computational cost is proportional to the variance, then
a simulation with a given $\tau$ and a target precision will cost, computationally, around an order of magnitude more in the absence of a Jastrow.
If this extra cost could be recovered by using time-steps ten times as large is something that would have to be checked on each system.

\subsection{Evaluation of ionization energies}\label{sec:IE}
Ionization energies (IEs) are typical quantities that are evaluated when new PPs are developed \revision{or tested.\cite{BFD, trail05_NCHF, trail05_SRHF, CEPP, CEPP-3d, eCEPP, ccECP, ccECP2, Krogel_OPT_PP_2016, Seth_QMC_1st_row_2011, Fahy_VMC_PP_1988, Ertekin_prb2013}} 
The $n$-th ionization energy IE$(n)$ is the energy necessary to remove one electron from an atomic (or molecular) species $X$ and charge $(n-1)$, {\em i.e.} to have $X^{n-1}\to X^n + \textrm{e}^{-}$.
Good PPs are expected to yield IEs estimations close to the corresponding all-electron (AE) evaluations.
Typically these checks are not performed at the QMC level, even for PPs specifically developed for QMC, but using Hartree-Fock, DFT,\footnote{
For instance, HF or DFT is used in the Burkatzki, Filippi, and Dolg's energy consistent pseudpotentials (BFD)\cite{BFD}, in the Trail and Needs' norm conserving Hartree-Fock pseudpotentials (TN-NC)\cite{trail05_NCHF} and smooth relativistic Hartree-Fock pseudpotentials (TN-DF)\cite{trail05_SRHF}.} 
or CCSD(T).\footnote{
For instance, CCSD(T) is used in Trail and Needs' correlated electron pseudopotentials (CEPP)\cite{CEPP, CEPP-3d} and the shape and energy consistent pseudopotentials\cite{eCEPP}, and in the Mitas and collaborators' correlation consistent effective core potentials (ccECP)\cite{ccECP, ccECP2}.}
%
Here, we test the FN-DMC evaluations using LA, TM, and DLA.

We considered the first three ionization energies of the carbon atom, and we performed calculations with the eCEPP pseudopotential\cite{eCEPP}, which has a He-core for the carbon atom.
These pseudopotentials perform very well at the CCSD(T) level of theory.
This can be seen in Table~\ref{tab:IE_C}, where
the absolute difference in the IE between AE and eCEPP evaluations is $<0.07$~eV,
and the relative difference is $< 0.5$\%.\footnote{
PPs an order of magnitude more accurate for IEs have recently been reported.\cite{ccECP}
}
\begin{table}[hbt]
\centering
\caption{
First three ionization energies of the carbon atom, in eV. We give experimental results (Exp.) and compare CCSD(T) results based on all electron (AE) and PP (eCEPP\cite{eCEPP}).
}\label{tab:IE_C}
\begin{tabular}{l  c c c c }
\hline
	&	Exp.	& AE\footnotemark[1] & eCEPP\footnotemark[2] & $\Delta_\textrm{eCEPP-AE}$ \\
\hline
IE(1) &	11.26 &	11.26 &	11.22 &	-0.04 \\
IE(2) &	24.38 &	24.36 &	24.29 &	-0.07 \\
IE(3) &	47.89 &	47.86 &	47.90 &	0.04 \\
\hline
\end{tabular}
\footnotetext[1]{Performed with Orca [48], using an aug-cc-pV(T,Q)Z basis set.}
\footnotetext[2]{Performed with Orca [48], using the aug-cc-pV5Z-TN basis.\cite{eCEPP} Differences w.r.t. aug-cc-pVQZ-TN are below 0.01 eV.}
\end{table}

\begin{figure*}[htb]
\centerline{
\includegraphics[height=3.0in]{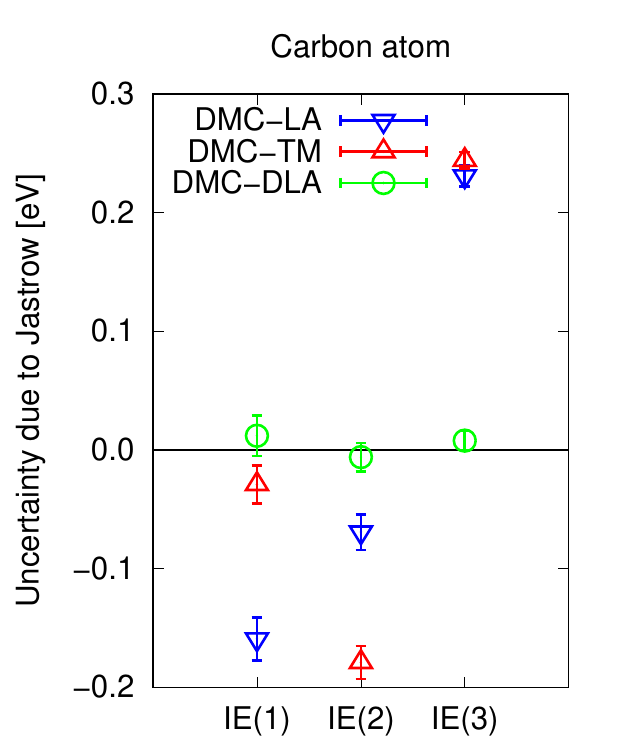}
}
\caption{\label{fig:dIE}
Uncertainty of the FN-DMC estimation of the $n$-th ionization energy IE($n$) of the carbon atom  due to the Jastrow, using an eCEPP pseudopotential\cite{eCEPP} and either LA, TM, or DLA. 
Uncertainty is evaluated as the difference between the estimation using two different sets of Slater-Jastrow wave functions having the same $\D$ and different $\J$ (see text).
Error bars correspond to a standard deviation due to the DMC sampling.
}
\end{figure*}

When PPs are deployed in FN-DMC the difference from the AE results can be much larger than what is found for CCSD(T) because of the localization error.
As usual, the localization error depends on the trial wave function.
To have an idea of the magnitude of the localization error, we performed two sets of FN-DMC calculations, both with eCEPP and a Slater-Jastrow function with the Slater determinant obtained from a DFT/LDA calculation.\footnote{
DFT/LDA calculation performed using the {\sc Orca} package\cite{orca}, which uses localized basis sets.
Reported FN-DMC results are converged in the DMC time-step and the size of the basis set.
}
The difference among the two sets is the Jastrow factor, that in one case (Ju) includes only e-e terms, 
and in the other case (Juxf) includes e-e, e-n and e-e-n terms.
The parameters of the Jastrow factors have been optimized for each atomic ion considered.
We have calculated the difference between the FN-DMC evaluations of IE$(n)$ in two sets, for LA, TM, and DLA. 
Notice that the difference between the corresponding Juxf and the Ju wave functions is solely due to the localization error, because the same $\D$ is used for Ju and Juxf, so the nodal surfaces are the same.
Results are reported on Figure~\ref{fig:dIE}.
By construction, DLA is not affected by the Jastrow uncertainty, whereas the uncertainty on both TM and LA is larger than 0.2~eV.
So, we see in this case that in LA and TM the choice of the Jastrow yields a localization error that can be more than two times larger than the CCSD(T) error of the eCEPP pseudopotential.
Given this large uncertainty in LA and TM, 
it is unclear what the most suitable Jastrow is.
Juxf is variationally better than Ju (smaller VMC energy and variance).
However, in Appendix~\ref{App:IE_locerr} we compare the difference between the PPs and the AE results, and it is unclear if Ju or Juxf is better.\footnote{\revision{Note that here better means a better error cancellation.}}
DLA solves the issue, because results from Ju and Juxf are equivalent. 

In which way can we improve the DMC accuracy with DLA? 
It can be improved systematically by changing the determinant part, for instance using a multi-determinant $\D$ term obtained using a method like the complete active space self consistent field (CASSCF),\cite{Valsson:2010p25419, Zimmerman:2009hh, Toulouse:2008p27527} the Configuration Interaction using a Perturbative Selection made Iteratively (CIPSI),\cite{CIPSI13, Scemama:2018} or the Antisymmetrized Geminal Power.\cite{JAGP_Casula2003, JAGP_diradicals_Zen2014}
DLA appears convenient in this case because we can anticipate improvements in the nodes and in the wave function, and we do not need to be concerned about the unpredictable effects of the Jastrow factor. 

\subsection{DLA yields stable DMC simulations}\label{sec:DLAstability}

Stability is a very important practical aspect of DMC simulations.
Instability in DMC is usually correlated with the quality of the trial wave function and of the pseudopotentials.
\revision{In particular, a possible issue in the trial wave functions, which can generate instability, is the generation of the determinant part $\D$ of the trial wave function via plane-wave DFT packages using suboptimal Kleinman-Bylander projectors.} 
In QMC (or in DFT calculations using localized basis) the non-local terms in the pseudopotential are evaluated using spherical harmonic projectors.
In contrast, in plane-wave DFT calculations the non-local terms in the pseudopotential are evaluated using Kleinman-Bylander projectors\cite{KleinmanBylander1982} (i.e. projected into pseudo-atom wave functions).
Thus, there is a possible inconsistency between the projection of the non-local terms at the DFT and DMC levels.
The inconsistency is even worse if the Kleinman-Bylander projectors are not obtained with the same DFT functional used in the DFT preparation of $\D$.
As a consequence, the method adopted to deal with the non-local part of the pseudopotential (LA, TM, or DLA) has a big impact on the stability of the DMC simulation.
%
We have observed that DLA is not affected by the instability issues of LA, and in all test simulations is as stable as TM and more stable than LA.


To illustrate this point, we consider carbon dioxide, CO$_2$, 
because in a previous study\cite{zen-pnas18} we noticed that the CO$_2$ molecule often leads to unstable DMC simulations.
We consider here the case of the monomer and dimer of CO$_2$ (configurations taken from Ref.~\onlinecite{zen-pnas18}) 
and we use CEPP pseudopotentials\cite{CEPP} for both oxygen and carbon. 
We obtain $\D$ from a DFT/LDA calculation using the {\sc PWSCF} code\cite{espresso}, with a 600~Ry  plane wave cutoff, and the molecular orbitals obtained were converted into splines\cite{splines-alfe04}. 
The Kleinman-Bylander projectors used in DFT are from Ref.~\onlinecite{CEPP}, and 
they were obtained from DFT/PBE so there is an inconsistency with the employed DFT functional.\footnote{
In this case the inconsistency among the Kleinman-Bylander projectors and the DFT functional used is likely the source of the instability of DMC simulations. 
Indeed, we can make the system much more stable by generating Kleinman-Bylander projectors using DFT/LDA atomic orbitals, for instance by using the {\sc ld1} code included in the {\sc Quantum Espresso} package.\cite{espresso}
However, in other systems it can be harder to improve the wave function and enhance stability.
Here we are intentionally considering a case that amplifies instability issues.
}
In QMC, we used a Jastrow factor $\J$ with e-e, e-n and e-e-n terms, 
and parameters were optimized by minimizing the \revision{local energy variance in VMC}  
yielding $\Var[E_L]_\text{VMC} = 0.956(2)$~au in the molecule.
The same identical trial wave function (i.e., no other optimization of $\J$)
has a much smaller \revision{local energy variance in VMC} if the DLA local energy $E_L^{DLA}$ (as obtained from the Hamiltonian given in Eq.~\ref{eq:H-DLAval}) is used:
$\Var[E_L^{DLA}]_\text{VMC} = 0.685(5)$~au.
Thus, the DLA Hamiltonian 
might already have advantages at the VMC level.

\begin{figure*}[htb]
\centerline{
\includegraphics[width=3.3in]{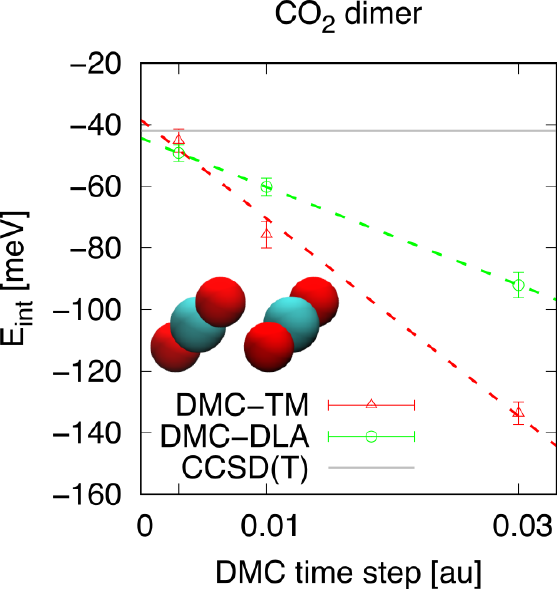}
}
\caption{\label{fig:Eb_CO2}
Carbon dioxide dimer configuration (inset) and interaction energy $E_\text{int}$ 
obtained from FN-DMC calculations and plotted as a function of the DMC time-step $\tau$. 
Dashed lines are obtained from a linear fit, and the extrapolated $\tau\to 0$ values are 
-38(5) meV and -44(1) meV for TM and DLA, respectively.
The reference CCSD(T) value is -42 meV.\cite{zen-pnas18}
DMC-LA is not reported because the simulations are unstable.
}
\end{figure*}

At the DMC level  
LA simulations are not possible, 
as population explosions happens so frequently that it was not possible to finish the DMC equilibration.\footnote{
It should be mentioned that some modifications to the Green's function can enhance slightly the stability.
For instance, within the ZSGMA scheme\cite{ZSGMA} to cure the local energy and the drift velocity divergences, it is possible to perform some DMC-LA simulations with large time-steps (e.g., $\tau>0.03$~au), 
but unfortunately
the time-step bias is too large.
}
In contrast, both TM and DLA yield stable DMC simulations at all the attempted time-steps (0.03, 0.01 and 0.003~au).
Moreover, the results of the DMC simulations seem reasonable despite the issue in the wave function.
The most interesting quantity to consider is the interaction energy $E_\text{int}$.
In this case, as the two carbon dioxide molecules are identical, the interaction can be evaluated as
$E_\text{int} = E_\text{dimer} - 2*E_\text{monomer}$, the difference between the energy of the dimer and twice the energy of the monomer.
The FN-DMC results with TM and DLA are reported in Figure~\ref{fig:Eb_CO2}.
Both methods are in good agreement with the reference CCSD(T) evaluation\cite{zen-pnas18} when we consider the infinitesimal time-step limit.
DMC-DLA is less affected by finite time-step bias than DMC-TM.
This is probably a consequence of the fact that the local energy variance in DLA is smaller than in TM.


%

\subsection{Good efficiency for DLA}\label{sec:DLAefficiency}

Efficiency is a fundamental property of a computational method.
DLA appears to be more efficient than LA and TM.
The efficiency of DMC can be estimated, as detailed in Appendix~\ref{App:costDMC}.
The choice among LA, TM, and DLA influences two quantities affecting the DMC efficiency:
the computational cost $T_\text{step}$ for a single DMC time-step,
and the variance $\sigma_\text{sys}^2$ of the local energy.
The most efficient methods will have smaller  $T_\text{step}$ and $\sigma_\text{sys}^2$.
DLA satisfies both the conditions.
A detailed comparison of the efficiency of LA, TM, and DLA is provided in Appendix~\ref{App:costDMC}.
The outcome is that DLA is more efficient in most of the simulations involving organic molecules by roughly 30\%.

\section{Conclusion}\label{sec:conclusion}

In this paper we have illustrated some drawbacks of FN-DMC in the presence of pseudopotentials.
Specifically:
(i) 
They generate unpredictable differences on results as a consequence of some arbitrary choices on the Jastrow functional form and the stochastic optimization procedure;
(ii) 
They might affect the reproducibility of results if different QMC packages are used;
(iii) 
The accuracy deteriorates whenever the Jastrow factor is not good, which is not easy to establish.
These issues are particularly problematic for precisely where FN-DMC is most needed and offers most promise.
We have shown that these issues arise essentially because the pseudopotentials have non-local terms.
Within both LA and TM, the projection scheme is affected by a subtle interaction between the Jastrow factor and the pseudopotentials.
In this paper we have introduced a new alternative approximation, called the DLA.
When FN-DMC deploys DLA the projected wave function and the associated energy are not affected by the Jastrow factor.
This solves the mentioned drawbacks.
The advantages of DLA have been illustrated on a few examples, including the evaluation of an interaction energy and ionization energies.
Moreover, the proposed algorithm appears as stable as TM and is much more stable than LA.
In terms of efficiency, DLA performs better than both LA and TM.

An interesting perspective for DLA is that it allows the development of general purpose Jastrow factors, which do not need a system dependent optimization and yield a validated accuracy.\footnote{\revision{
It sould be mentioned that there have been previous attempts to use wave functions that do not need optimization in QMC,\cite{plasma_Ceperley_PRB1978, backflow_Holzmann_PRE2003, plasmon_Wood_JPCM2006}
although they have focused on relatively simpler systems and do not deal with atomic PPs and localization errors in DMC.}}
In this way, DLA opens the way to the automation of the FN-DMC, which can make DMC easier and less labour intense to use.

The DLA method is already implemented in the CASINO\cite{casino}, TurboRVB\cite{TurboRVB}, 
\revision{ and QMCPACK\cite{QMCPACK}} packages.


\begin{acknowledgements}
\revision{ We thank Sandro Sorella, Pablo L\'{o}pez R\'{i}os and Paul Kent for useful discussions on the presented methodology.
Moreover, we acknowledge the prompt implementation of DLA in TurboRVB by Sandro Sorella, and in QMCPACK by Paul Kent and Ye Luo. We are grateful to Pablo L\'{o}pez R\'{i}os for the help in implementing DLA in the CASINO package.}
A.Z. and D.A. are supported by the Air Force Office of Scientific Research, Air Force Material Command, US Air Force, under Grant FA9550-19-1-7007. 
A.Z. and A.M. were also supported by the European Research Council (ERC) under the European Union's Seventh Framework Program (FP/2007-2013)/ERC Grant Agreement 616121 (HeteroIce project).
J.G.B acknowledges support by the Alexander von Humboldt foundation.
We are also grateful, for computational resources, 
to ARCHER UK National Supercomputing Service, United Kingdom Car-Parrinello (UKCP) consortium (EP/F036884/1), 
the London Centre for Nanotechnology, University College London (UCL) Research Computing, 
Oak Ridge Leadership Computing Facility (DE-AC05-00OR22725), 
and the UK Materials and Molecular Modelling Hub, which is partially funded by EPSRC (EP/P020194/1).
\end{acknowledgements}

\appendix

\section{DMC implementation and Fixed-Node approximation}\label{app:DMCimplementation}

In DMC the non-negative and normalized function $f(\R,t)$ is interpreted as a probability density distribution, 
which is represented at each time $t$ via a large number of electronic configurations $\R_i(t)$, also called walkers, and their associated weights $w_i(t)$.
Walkers $\R_i(t)$ and weights $w_i(t)$ evolve in time according to a process that is ultimately determined by the Green's function.\cite{foulkes01} 
A key issue is that the Green's function $G(\R' \leftarrow \R,t)$ in eq. (\ref{eq:Gdef}) 
does not impose any anti-symmetry constraint to the system, so the ground state $\Phi$ obtained according to the imaginary time projection in eq. (\ref{eq:DMCprojection}) would be bosonic, as it has a lower energy than the corresponding fermionic system.
The traditional and most effective way to impose anti-symmetry in $\Phi$ is to adopt the {\bf fixed node} (FN) approximation: walkers are not allowed to cross the nodal surface of the trial wave function $\PsiTR$. 
In other terms, the Hamiltonian $\hat H$ is replaced by the FN Hamiltonian 
$\hat H_{FN} \equiv \hat H + \hat V_{FN} \,, $ 
where $\hat V_{FN}$ is an infinite wall at the nodal surface of $\Psi_T$.
The projected function $\Phi_{FN}(\R)$ obtained in this way has the same nodes as $\PsiTR$, 
and it is the exact solution for the Hamiltonian $\hat H_{FN}$, namely 
$\hat H_{FN} \left| \Phi_{FN} \right> = E_{FN} \left| \Phi_{FN} \right>$.
This implies that 
$E_{FN}= {\left< \Phi_{FN} \right| \hat H_{FN} \left| \Psi_T \right> \over \left< \Phi_{FN} \middle| \Psi_T \right>}$,
and noticing that  $\hat H_{FN} \left| \Psi_T \right> = \hat H \left| \Psi_T \right>$ because $\hat V_{FN}\ne 0$ only in the nodal surface of $\Psi_T$,
so we can evaluate $E_{FN}$ using the mixed estimator:
\begin{equation}\label{eq:eFN_mix}
E_{FN} =
{ \int \Phi_{FN}(\R) \PsiTR E_L(\R) d\R \over \int \Phi_{FN}(\R) \PsiTR d\R } 
\,.
\end{equation}
The FN energy $E_{FN}$ is an upper bound to the exact energy $E_0$, 
namely $E_{FN} \ge E_{0}\,,$
because $\hat H_{FN} \ge \hat H$.
The approach is exact if the nodes of $\PsiTR$ are exact, because in this case $\hat H_{FN} \left| \Phi_{FN} \right> = \hat H \left| \Phi_{FN} \right>$.
Thus, the quality of the FN approximation is determined by the quality of the nodes of the trial function $\PsiTR$.

\section{ Hamiltonian for all-electrons versus Hamiltonian for valence electrons and pseudopotentials}\label{app:AEvsPP}

The electronic Hamiltonian in Born-Oppenheimer approximation is:
\begin{equation}\label{eq:Hae}
\hat H = 
	- {1\over 2} \sum_i^{N} \nabla_i^2 
	+ \sum_{i<j}^{N} {1 \over r_{ij}} 
	- \sum_i^{N} \sum_\alpha^{M} {Z_\alpha \over r_{i\alpha}}
\end{equation}
where
roman letters are used to label the $N$ electrons and greek letters to label the $M$ ions; 
$r_{ij}=| {\bf r}_i - {\bf r}_j |$ is the distance between electrons $i$ and $j$;
$r_{i \alpha}=| {\bf r}_i - {\bf r}_\alpha |$ is the distance between electron $i$ and ion $\alpha$.
The first term in the right hand size is the kinetic energy $\hat K$ and the other two terms define a local potential $\hat V_L$. 

The electrons can be separated into $N_c$ core and $N_v$ valence electrons, such that
$N=N_v+N_c$.
Thus, equation (\ref{eq:Hae}) can be recasted to:
\begin{eqnarray}\label{eq:Hsplit}
\hat H &=& \hat H_{v} + \hat H_{c} + \hat H_{cv} 
	\\
\hat H_{v} &=& 	
	- {1\over 2} \sum_i^{N_v} \nabla_i^2 
	+ \sum_{i<j}^{N_v} {1 \over r_{ij}} 
	- \sum_i^{N_v} \sum_\alpha^{M} {Z_\alpha \over r_{i\alpha}}
	\\
\hat H_{c} &=&
	- {1\over 2} \sum_i^{N_c} \nabla_i^2 
	+ \sum_{i<j}^{N_c} {1 \over r_{ij}} 
	- \sum_i^{N_c} \sum_\alpha^{M} {Z_\alpha \over r_{i\alpha}} 
	\\
\hat H_{cv} &=&
	+ \sum_{i}^{N_c} \sum_{j}^{N_v} {1 \over r_{ij}} 
\end{eqnarray}
where $\hat H_{v}$ and $\hat H_{c}$ are the components of the Hamiltonian involving only the valence electrons and the core electrons, respectively, and  
$\hat H_{cv}$ is the explicit pairing interaction between the core and the valence electrons.

All electron calculations in QMC are computationally very expensive, having a scaling roughly proportional to $Z_\alpha^6$,\cite{Ceperley_Zscaling_1986, Hammond_Zscaling_1987, Needs_Zscaling_2005} 
because close to the nucleus the local energy has large fluctuations, yielding a large variance and thus requiring very small time-steps.
As many properties of interest are determined by the behavior of the valence electrons, it is often convenient to use pseudopotentials to represent the core electrons, especially in heavy nuclei.
The Hamiltonian for calculations with pseudopotentials is then of the following:
\begin{equation}\label{eq:Hpp}
\hat H_\text{PP} = 
	- {1\over 2} \sum_i^{N_v} \nabla_i^2 
	+ \sum_{i<j}^{N_v} {1 \over r_{ij}} 
	+ \sum_i^{N_v} \sum_\alpha^{M} \hat V_{\text{PP}}^\alpha ({\bf r}_i)
\end{equation}
where the first two terms in the RHS are the kinetic and electron-electron interactions, respectively, as expressed also in $\hat H_v$.
The interaction between electron $i$ and ion $\alpha$ are described as follows:
\begin{equation}\label{eq:Vpp}
\hat V_{\text{PP}}^\alpha ({\bf r}_i) = 
V_\text{loc}^\alpha(r_{i\alpha}) + 
\sum_{l=0}^{l_\text{max}} V_{l}^\alpha(r_{i\alpha}) 
\hat P_{l}^\alpha
\end{equation}
where $V_\text{loc}^\alpha(r_{i\alpha})$ is the local part of the pseudopotential of ion $\alpha$, 
and $V_{l}^\alpha(r_{i\alpha})$ are the non-local components, which are applied via the projector:
$$
\hat P_{l}^\alpha = \sum_{m=-l}^{+l} \left| Y^\alpha_{l,m} \middle> \middle< Y^\alpha_{l,m} \right|
$$
where $\left| Y^\alpha_{l,m} \right>$ are spherical harmonics centered on nucleus $\alpha$. 
The idea behind $\hat V_{\text{PP}}^\alpha ({\bf r}_i)$ is that it represents an effective potential that reproduces the effects of both the nucleus and the core electrons on the valence electrons. 
However, there is not an exact mapping, or a thermodynamic integration, providing the pseudopotentials.
Indeed some criteria needs to be chosen to produce them, and they need to be tested at some level of theory. 
Moreover, it has to be noticed that in independent-electron theories, such as HF and DFT, the separation of the electrons among core and valence can be in principle exact, while it cannot be exact in QMC or in other many-body approaches, because of the electronic correlation.
Thus, although the employment of pseudopotentials in QMC is most of the times necessary for efficiency reasons, it 
can yield errors which cannot be easily quantified. 
\revision{
Accurate pseudopotentials for QMC are so far available only for a fraction of the periodic table.
The few cases where QMC proves less accurate than DFT are typically related with poor quality PPs used in QMC.\cite{Saritas_jctc2017}
A crucial property of a PP is its transferability, which is affected by the range of the non-local potential terms (the smaller the better) and the inclusion of higher angular momentum channels.\cite{Tipton_highLpseudo_prb2014}
}

\section{ Jastrow interaction with pseudopotential non-local term}\label{app:JvsPP}

In sections \ref{sec:LA} and \ref{sec:TM} we have shown that the non-local operators needs to be localized using the trial wave function $\PsiT$,
such that the localized non-local operator ${\hat V_{NL} \PsiT \over \PsiT}$ acts on a generic function $ \xi(\R)$ as follows:
$$
{\hat V_{NL} \PsiT \over \PsiT} \xi(\R) = {\cal V}_{NL}^{loc}(\R) \xi(\R) \,.
$$
%
By using the addition theorem for spherical harmonics and equations \ref{eq:Hpp} and \ref{eq:Vpp}, it can be shown that ${\cal V}_{NL}^{loc}(\R)$ can be evaluated as follows:
\begin{equation}\label{eq:Vncloc}
{\cal V}_{NL}^{loc}(\R) = \sum_i^{N_v} \sum_\alpha^M \sum_{l=0}^{l_\text{max}} 
{2l+1\over4\pi} V_l^\alpha(r_{i\alpha}) 
\int P_l(\cos\theta'_{i\alpha}) {\PsiT({\bf r}_1,\ldots,{\bf r}_{i-1},{\bf r'}_{i},{\bf r}_{i+1}\ldots,{\bf r}_{N_v}) \over \PsiT({\bf r}_1,\ldots,{\bf r}_{N_v})} d\Omega'_{i\alpha}
\end{equation}
where 
the angular integration is over the sphere passing through the $i$-th electron and centered on the $\alpha$-th atom,
$P_l$ is the Legendre polynomial of degree $l$,
and $\cos\theta'_{i\alpha} = {{\bf r}'_{i\alpha} \cdot {\bf r}_{i\alpha}  \over r'_{i\alpha} r_{i\alpha} }$.

Therefore, in the evaluation of ${\cal V}_{NL}^{loc}(\R)$ we have to integrate the ratio between the trial wave functions for electrons displaced along spheres centered on every pseudo-atom.
By considering that  
$ \PsiT = \D * \eJ $,
it is clear that this ratio has to be taken for both the determinant part $\D$ and the Jastrow factor  $\eJ$.
This expression explains why both LA and TM yield to total energies that depend on the Jastrow factor: $\J$ effectively changes the potential term in the effective Hamiltonian, due to the localization of the non-local term. 

We could wonder 
what parts of the Jastrow factor give rise to this issue, and 
if it is possible to use a Jastrow not affecting the ratio in the RHS of equation \ref{eq:Vncloc}.
Although there are differences in the Jastrow factor implemented in the different codes, they typically can be expressed in the following way:
\begin{equation}
{\cal J}(\R) = 
\sum_{i<j} \delta_{\sigma_i,\sigma_j} u_{\uparrow\uparrow}(r_{ij}) + 
\sum_{i<j} (1-\delta_{\sigma_i,\sigma_j}) u_{\uparrow\downarrow}(r_{ij}) + 
\sum_\alpha \sum_i \chi_\alpha(r_{i\alpha}) + 
\sum_\alpha \sum_{i<j} f_\alpha(r_{i\alpha},r_{j\alpha},r_{ij})
\end{equation}
where $u_{\uparrow\uparrow}$ and $u_{\uparrow\downarrow}$ are homogeneous two-electron correlation terms, or e-e, describing the interation between like-spin and unlike-spin pairs, 
$\chi_\alpha$ is a e-n term depending on the distance of any electron from the $\alpha$-th atom, 
and $f_\alpha$ is a three-body term, or e-e-n, describing the interaction between electron pairs in proximity of the $\alpha$-th atom.
Different packages can provide different parametrization of these terms, and sometimes additional terms.\footnote{
For instance, in CASINO\cite{casino} Jastrow terms involving arbitrary numbers of particles can be introduced,\cite{LopezRios_gJastrow_2012} and in TurboRVB\cite{TurboRVB} there is also the possibility to use four-body e-e-n-n terms, and the e-e-n term is an explicit function of $({\bf r}_{i\alpha},{\bf r}_{j\alpha})$, see Ref.~\onlinecite{Zen_jctc2013}.
}
%
The important aspect is that in the displacement 
$\br_i \to \br'_i$
performed in the angular integration in equation \ref{eq:Vncloc}, we have that
$r_{i\alpha} \to r'_{i\alpha}=r_{i\alpha}$ 
but
$r_{ij} \to r'_{ij} \ne r_{ij}$.
Thus, the e-n term $\chi_\alpha$ is not responsible for the Jastrow dependence of ${\cal V}_{NL}^{loc}(\R)$,
while the e-e and e-e-n terms are at the source of the issue.
One could in principle decide to do not use the three-body term $f_\alpha$,
but the same cannot be done for the terms $u_{\uparrow\uparrow}$ and $u_{\uparrow\downarrow}$, as they are the most important in the Jastrow factor, responsible for most of the correlation captured by $\PsiT$ and the decrease in the variance of the local energy.\cite{gruneis_perspective_2017}

\section{How large are errors due to PPs in IE evaluations?}\label{App:IE_locerr}

Ideally, a perfect PP delivers exactly the same energy differences as AE calculations.
In FN-DMC the situation is complicated by the fact that localization errors appear when PPs are used.
In section~\ref{sec:IE} it is shown that these localization errors can be large, and that with LA and TM there is a big dependence on the Jastrow factor in the trial wave function.
For instance, it is not clear if both the wave functions Ju and Juxf discussed in section~\ref{sec:IE} show a sizable localization error, or if instead one wave function is responsible for most of the error.

\begin{figure*}[htb]
\centerline{
\includegraphics[height=3.0in]{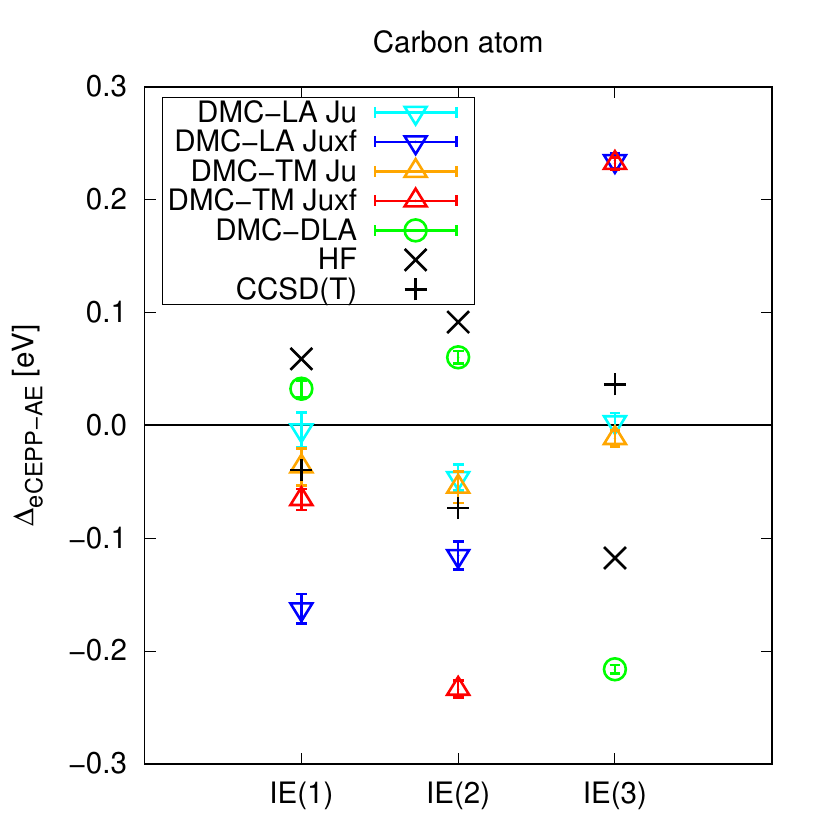}
}
\caption{\label{fig:dIE_PP-AE}
Difference $\Delta_\text{eCEPP-AE}$ between the IE$(n)$ evaluation with eCEPP PPs and with all-electrons (AE).
FN-DMC with PPs yields different results with different $\PsiT$, and 
we consider two sets of Slater-Jastrow functions, dubbed Ju and Juxf (see text). 
In all reported DMC calculations, $\D$ is a single Slater determinant obtained from DFT/LDA. 
Results using LA, TM, and DLA are shown. 
As a comparison, outcomes from Hartree Fock (HF) and CCSD(T) are also reported.
}
\end{figure*}

In order to investigate this, we have considered the difference $\Delta_\text{eCEPP-AE}$ between the eCEPP and the AE results, the IE$(n)$ evaluated with LA, TM and DLA.
In the AE calculation we have used the same level of theory to generate the wave function, so a Slater-Jastrow function with the determinant from DFT/LDA.\footnote{ The employed Jastrow for the AE calculations has e-e, e-n and e-e-n terms. Notice that in AE calculations there is no localization error, i.e., the Jastrow factor does not affect the accuracy and does not lead to any uncertainty.}
Results are reported in Figure~\ref{fig:dIE_PP-AE}.
%
The plot highlights some unexpected behavior.
We would have expected that the Juxf yields better results than Ju, 
because Juxf has more parameters and include Ju as a special case, and indeed it yields a lower variational energy and variance than Ju.
However, Juxf does not show a smaller $\Delta_\text{eCEPP-AE}$  compared to Ju in any of the IEs.
In fact, the worse wave function, Ju, generates the smallest $\Delta_\text{eCEPP-AE}$ in LA and TM.
On the other hand, DLA yields $\Delta_\text{eCEPP-AE}$ quite similar in absolute value to the best  LA or TM case, 
but in DLA the sign is inverted.
We notice a clear correlation between DLA and HF errors.
This correlation suggests that the $\Delta_\text{eCEPP-AE}$ in DLA might just reflect the limitations of the determinant part of the wave function.

\section{Estimation of computational cost of a FN-DMC simulation with LA, TM, and DLA}\label{App:costDMC}

The efficiency of a DMC calculation depends on the computational resources required to achieve a target stochastic error $\sigma_\text{target}$ on the quantity of interest.
In most cases we are interested in the energy, and in this case the computational time spent $T_\text{cost}$ is
\begin{equation}\label{eq:DMCcost}
T_\text{cost} = T_\text{step} * {t_\text{ac} \over \tau} \left( {t_\text{ac} \over t_\text{eq} } N_w + {\sigma_\text{sys}^2 \over \sigma_\text{target}^2 } \right) 
\,,
\end{equation}
where 
$T_\text{step}$ is the computational time of a single DMC step (which clearly depends on the specific architecture where the calculation is performed), 
$t_\text{ac}$ and $t_\text{eq}$ are the autocorrelation and equilibration times (which are roughly the same and typically of the order of 1~au),
$N_w$ is the number of walkers,
and 
$\sigma_\text{sys}^2$ is the variance of the local energy in the corresponding DMC scheme.\footnote{
For a derivation of eq.~(\ref{eq:DMCcost}) see the SI of Ref.~\onlinecite{zen-pnas18}.
} 
The first term into the parenthesis is due to the equilibration time in DMC, which has to be removed from the sampling, and that typically is negligible compared to the second term into the parenthesis, which instead comes from the statistical sampling.
The quantities on the RHS of eq. (\ref{eq:DMCcost}) that are affected by the choice of LA, TM or DLA are only 
$T_\text{step}$ and $\sigma_\text{sys}^2$.
In particular, $\sigma_\text{sys}^2$ is 
$\Var[E_L]_\text{DMC-LA}$,
$\Var[E_L]_\text{DMC-TM}$,
and
$\Var[E_L^{DLA}]_\text{DMC-DLA}$
for LA, TM, and DLA, respectively.

The dependence on $T_\text{step}$ is straightforward:
if we name $T_\text{step}^{LA}$, $T_\text{step}^{TM}$ and $T_\text{step}^{DLA}$ the cost per step in the LA, TM and DLA approaches, respectively, then 
$$
T_\text{step}^{DLA} < T_\text{step}^{LA} < T_\text{step}^{TM} \,.
$$
Indeed, in TM the operations are just the same performed in LA, plus those to perform the moves connected to the Green's function $T_{NL}(\R' \leftarrow\R)$ (T-moves).
On the other hand, in DLA there are fewer operations than in LA because the Jastrow factor does not need to be considered when evaluating the non-local part of the pseudopotential (see details in Appendix~\ref{app:JvsPP}).
However, the observed difference in the costs of LA, TM and DLA is relatively small (just a few percent).

More important for the efficiency is the dependence on the variance $\sigma_\text{sys}^2$.
First of all, we have to notice that here the variance under consideration is the one relative to the corresponding DMC sampling, which is
$\Var[E_L]_\text{DMC-LA}$,
$\Var[E_L]_\text{DMC-TM}$,
and
$\Var[E_L^{DLA}]_\text{DMC-DLA}$
for LA, TM and DLA, respectively.
The LA and TM approaches are strictly related with the variational variance
$\Var[E_L]_\text{VMC}$;
the local energy is calculated in the same way but the underlying probability distribution is different,
and in particular different values of the time-step change the sampling and the corresponding variance.
In DMC-LA with large $\tau$ the variance becomes the same as the variational variance (this is likely a consequence of the Metropolis step to enforce detailed balance after the drift-diffusion step), 
while at small $\tau$ we notice that the DMC-LA variance is typically slightly smaller than the VMC variance.
In DMC-TM the variance converges to the same variance of DMC-LA for small $\tau$, but for large $\tau$ it is larger than the VMC and the DMC-LA variance.\footnote{
This is likely because the T-moves are performed after the Metropolis step, so the TM algorithm does not satisfy detailed balance.
}
In DMC-DLA the variance has roughly the same relation with the variational DLA variance that 
the DMC-LA variance has with the VMC variance.
Thus, the difference in the variance between DMC-LA and DMC-DLA is mostly captured by the difference in the corresponding variational variances,  
$\Var[E_L]_\text{VMC}$ and
$\Var[E_L^{DLA}]_\text{VMC}$.
Notice that the VMC sampling is precisely the same if $\PsiT$ is the same, and the difference only comes from the Hamiltonian, thus the local energy.

\begin{figure}[htb]
\centerline{\includegraphics[width=3.3in]{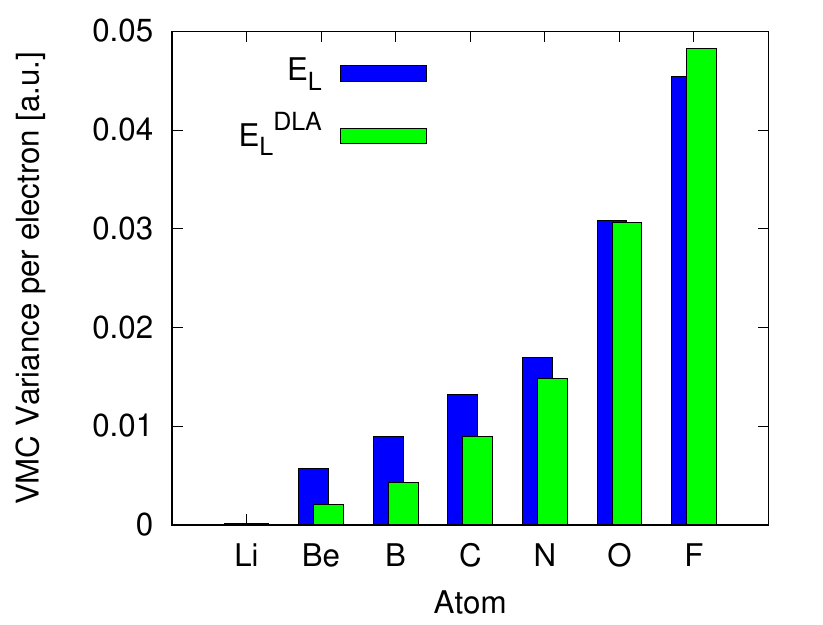}} 
\caption{\label{fig:VarAtoms}
\revision{Local energy variances in VMC}, $\Var[E_L]_\text{VMC}$ in blue and $\Var[E_L^{DLA}]_\text{VMC}$ in green, per valence electron on the first-row atoms, using eCEPP pseudopotentials\cite{eCEPP} (He-core), a Jastrow factor with 
e-e, e-n and e-e-n terms,  
and a single Slater determinant obtained from a DFT/LDA calculation with the aug-cc-pVQZ-eCEPP basis.\cite{eCEPP}
}
\end{figure}
In order to investigate this difference, in Figure~\ref{fig:VarAtoms} we report the variational variances 
($\Var[E_L]_\text{VMC}$ and $\Var[E_L^{DLA}]_\text{VMC}$) 
per electron on the first-row atoms, using eCEPP pseudopotentials\cite{eCEPP}, which are He-core for all the reported atoms.
The $\D$ in $\PsiT$ was obtained from a DFT/LDA calculation performed with the {\sc Orca} package\cite{orca} and with the localized basis set aug-cc-pVQZ-eCEPP provided in Ref.~\onlinecite{eCEPP}.\footnote{
Using a localized basis we get rid of possible issues with the Kleinman-Bylander projectors in plane-wave DFT calculations. Moreover, we crosschecked in the carbon atom case that the $\D$ obtained from plane-wave DFT calculations with a 600~Ry cutoff and from open-system DFT calculations with an aug-cc-pVQZ-eCEPP basis have similar quality. Indeed, the variance obtained when using the same Jastrow factor is roughly the same.
}
The Jastrow factor used has e-e, e-n and e-e-n terms, with parameters optimized in any atom to minimize the variational variance.\footnote{
Here there is a possible ambiguity, as one could wonder if we have to minimize the variance $\Var[E_L]_\text{VMC}$ or $\Var[E_L^{DLA}]_\text{VMC}$, or provide different optimization if we intend to use the original Hamiltonian $\hat H$ or the DLA version $\hat H^{DLA}$.
However, in all cases considered we noticed that the parameters of $\J$ that minimize $\Var[E_L]_\text{VMC}$ are, within the stochastic uncertainty, the same as those that minimize $\Var[E_L^{DLA}]_\text{VMC}$, and vice-versa. 
}
Figure~\ref{fig:VarAtoms} shows that $\Var[E_L^{DLA}]_\text{VMC}$ is smaller than $\Var[E_L]_\text{VMC}$ for most elements with the exception of the fluorine atom where the DLA variance is +6\% larger. 
In particular, in the carbon atom the variance is -32\% smaller, in the nitrogen atom is -13\%, in oxygen atom in -0.5\%.

A natural question at this point is: which part of $\J$ is mostly involved in producing a difference between $E_L$ and $E_L^{DLA}$?
In Appendix~\ref{app:JvsPP} we show that the e-n term drops out when we evaluate the non-local potential term, thus it produces no difference between $E_L$ and $E_L^{DLA}$.
The terms to consider are thus the e-e and the e-e-n.
In Table~\ref{tab:var_B} we report the variational variances on the boron atom with different parametrizations  of the $\J$.
It shows than the e-e term produces most of the difference.
Moreover, one could wonder how much this difference is affected by the choice of the pseudopotentials.
In Table~\ref{tab:var_B} we provide the variance also for the ccECP pseudopotential\cite{ccECP}.
We notice that, despite in absolute terms eCEPP and ccECP yield different variances, in relative terms the difference between $E_L$ and $E_L^{DLA}$ is the same.

\begin{table}[hbt]
\centering
\caption{
Variational variance, in au, $\Var[E_L]_\text{VMC}$ and $\Var[E_L^{DLA}]_\text{VMC}$
for the boron atom,
simulated using 
He-core pseudopotentials, 
a single Slater determinant obtained from a DFT/LDA calculation
and different parametrizations of the Jastrow factor $\J$, each with parameters optimized to minimize the variance.
For the  eCEPP pseudopotentials\cite{eCEPP} we used the aug-cc-pVQZ-eCEPP basis provided by \citet{eCEPP},
while for the ccECP pseudopotentials\cite{ccECP} we used the QZ basis provided by \citet{ccECP}.
}\label{tab:var_B}
\begin{tabular}{l  c c c }
\hline
Pseudo & Jastrow	&  $\Var[E_L]_\text{VMC}$ & $\Var[E_L^{DLA}]_\text{VMC}$ \\
\hline
eCEPP &	no	&	0.181(3) 	& 0.181(3) \\
eCEPP &	e-e	&	0.0401(1)	& 0.0317(1) \\
eCEPP &	e-e, e-n &	0.0285(3) & 0.0135(1) \\
eCEPP &	e-e, e-n, e-e-n &	0.0278(2) & 0.0133(1) \\
\hline
ccECP &	e-e, e-n, e-e-n &	0.0355(4) & 0.0166(1) \\
\hline
\vspace{-0.4cm}
\end{tabular}
\end{table}

The lower variance automatically translates into a smaller stochastic error in the DMC evaluations with same sampling.
Thus, DLA is more efficient in most of the simulations involving organic molecules, because they are characterized by the presence of many carbon atoms, where DLA is roughly 30\% faster than LA and even more compared to TM (where it crucially depends on the time-step value).


%

\end{document}